\documentstyle[12pt,epsf,axodraw]{article}
\begin{document}
\begin{titlepage}
\hskip 11cm \vbox{
\hbox{BUDKERINP/99-62}
\hbox{UNICAL-TH 99/4}
\hbox{July 1999}}
\vskip 0.3cm
\centerline{\bf THE QUARK IMPACT FACTORS$^{~\ast}$}
\vskip 0.8cm
\centerline{  V.S. Fadin$^{a~\dagger}$, R. Fiore$^{b~\ddagger}$,
M.I. Kotsky$^{a~\dagger}$, A. Papa$^{b~\ddagger}$}
\vskip .3cm
\centerline{\sl $^{a}$ Budker Institute for Nuclear Physics and Novosibirsk}
\centerline{\sl State University, 630090 Novosibirsk, Russia}
\centerline{\sl $^{b}$ Dipartimento di Fisica, Universit\`a della Calabria
and Istituto}
\centerline{\sl Nazionale di Fisica Nucleare, Gruppo collegato di Cosenza,}
\centerline{\sl Arcavacata di Rende, I-87036 Cosenza, Italy}
\vskip 1cm
\begin{abstract}
We calculate in the next-to-leading approximation the non-forward quark 
impact factors for both singlet and octet color representation in the 
$t$-channel. The integral representation of the octet impact 
factor in the general case of arbitrary space-time dimension and massive quark 
flavors is used to check the so-called "second bootstrap condition" for the 
gluon Reggeization at the next-to-leading logarithmic approximation in 
perturbative QCD. We find that it is satisfied for both helicity conserving 
and non-conserving parts. The integrations are then performed
for the explicit calculation of the impact factors in the massless quark case.
\end{abstract}
\vfill

\hrule
\vskip.3cm
\noindent
$^{\ast}${\it Work supported in part by the Ministero italiano
dell'Universit\`a e della Ricerca Scientifica e Tecnologica, in part
by INTAS and in part by the Russian Fund of Basic Researches.}
\vfill
$ \begin{array}{ll}
^{\dagger}\mbox{{\it e-mail address:}} &
 \mbox{FADIN, KOTSKY ~@INP.NSK.SU}\\
\end{array}
$

$ \begin{array}{ll}
^{\ddagger}\mbox{{\it e-mail address:}} &
  \mbox{FIORE, PAPA ~@FIS.UNICAL.IT}
\end{array}
$
\vfill
\vskip .1cm
\vfill
\end{titlepage}
\eject

\section{Introduction}
\setcounter{equation}{0}

The BFKL equation~\cite{BFKL} is very important for the theory of the 
Regge processes at high energy $\sqrt s$ in perturbative QCD. In particular,
it 
can be used together with the DGLAP equation~\cite{DGLAP} for the description 
of the structure functions for the deep inelastic $e p$ scattering at small 
values of the Bjorken variable $x$. It was derived more than twenty years 
ago in the leading logarithmic approximation (LLA) for the Regge region~\cite{BFKL},
that means summation of all the terms of the type $(\alpha_s \ln s)^n$.
Recently also the radiative corrections to the kernel of the equation
have been calculated~\cite{FL89}-\cite{FFFK} and the explicit form of the kernel 
in the next-to-leading approximation (NLA) for the case of forward scattering was 
found~\cite{FL98,CC98}. A number of subsequent papers (see, for instance~\cite{NLA})
were devoted to the investigation of its consequences.

In the BFKL approach the high energy scattering amplitudes are expressed 
in terms of the Green function of two interacting Reggeized gluons and of 
the impact factors of the colliding particles~\cite{BFKL,FL98,rio98,dis98}. 
The BFKL equation allows to determine this Green function for forward scattering
($t=0$ and singlet color state in the $t$-channel). The impact factors
must be calculated separately and only in some cases (such as strongly-virtual 
photon or hard mesons) perturbation theory is applicable.  

The key role in the derivation of the BFKL equation is played by the gluon 
Reggeization. In the LLA the Reggeization was noticed in the first orders of 
the perturbation theory and, assuming that it is correct to all orders, the
equation for the $t$-channel partial waves of the elastic scattering amplitudes
was derived~\cite{BFKL}. It is evident that, for gluon quantum numbers in the
$t$-channel, the solution of this equation must reproduce the gluon
Reggeization. 
This was explicitly demonstrated in Ref.~\cite{BFKL}.
This ``bootstrap'' condition represents a stringent test of the gluon 
Reggeization, although it cannot be considered as a proof. In the LLA 
such proof was constructed in Ref.~\cite{BFL79}. In the NLA the gluon
Reggeization has only been checked in the first three orders of the
perturbation theory~\cite{F}. Since it forms the base of the BFKL approach,
a more stringent test is desirable. As well as in the LLA, such test is
provided by the ``bootstrap'' condition.
Using the gluon Reggeization as a base, it is possible to represent the 
high energy scattering amplitudes as a convolution of the impact factors of 
the colliding particles and of the Green function for two interacting 
Reggeized gluons (see Eq.~(\ref{ImA}) below) not only for the forward
scattering, but also for non-zero momentum transfer $\sqrt{-t}$ and arbitrary 
color states in the $t$-channel~\cite{FF98}. 
For the case of octet color representation, the requirement
of self-consistency leads to two ``bootstrap'' equations for the gluon
Reggeization 
in the NLA (Eqs.~(34) and~(35) in Ref.~\cite{FF98}). Besides providing a
stringent check of the gluon Reggeization, these equations are important since they
contain almost all the values appearing in the NLA BFKL kernel and so provide 
a global test of calculations carried out over a long period of 
time~\cite{FL89}-\cite{FFFK} and only 
in a small part independently performed~\cite{CCH} or checked~\cite{BRN98,DS98}.

The first bootstrap condition involves the kernel of the non-forward
BFKL equation for octet color representation in the $t$-channel,
expressed in terms of the effective vertices for the Reggeon-Reggeon
interaction.
The explicit demonstration that it is satisfied in the part concerning the 
quark--anti-quark contribution was given in Ref.~\cite{FFP98} for arbitrary
space-time dimension. The second bootstrap condition involves the impact
factors 
of the scattered particles for octet color representation in the $t$-channel,
expressed in terms of the effective vertices for the Reggeon-particle
interaction.
For the case of gluon impact factors, the explicit proof that this equation is
satisfied has been given in Ref.~\cite{FFKP99} for arbitrary space-time
dimension
and massive quarks for both the helicity conserving and non-conserving parts.
It must be stressed that for this proof it is sufficient to use the 
integral representation of the NLA non-forward gluon impact factors with
color octet states in the $t$-channel and it is not necessary to perform 
explicit integration and to consider other color states. However, the impact 
factors have their own value, therefore in Ref.~\cite{FFKP99} they
have been obtained in the gluon case for arbitrary color representation
in the $t$-channel and the explicit integrations have been explicitly carried 
out in the massless quark case. 

The main aim of this paper is to demonstrate the fulfillment of the second 
bootstrap condition also in the case of quark impact factors, along the same
lines
as in Ref.~\cite{FFKP99}. We determine an integral representation of the 
NLA non-forward quark impact factors for arbitrary space-time dimension and 
representation of the color group and check that the second bootstrap
condition for the octet representation is satisfied for both the helicity 
conserving and non-conserving parts. The integrations are then carried out 
explicitly in the case of massless quarks.

The paper is organized as follows: in Section~2 we explain the method of 
calculation, in Section~3 and~4 we obtain the integral representation of the
contributions to the quark impact factors from one-quark and from quark-gluon 
intermediate states, respectively; in Section~5 the check of the second
bootstrap 
condition is explicitly demonstrated for both the helicity conserving and 
non-conserving parts; in Section~6 the integrations representing the 
NLA non-forward quark impact factors are carried out in the massless quark
case. Section~7 contains the summary and a short discussion.

\begin{figure}[tb]      
\begin{center}
\begin{picture}(240,200)(0,0)

\ArrowLine(0,190)(75,190)
\ArrowLine(165,190)(240,190)
\Text(37.5,200)[]{$p_A$}
\Text(202.5,200)[]{$p_{A'}$}
\Text(120,190)[]{$\Phi_{A'A}(\vec q_1,\vec q)$}
\Oval(120,190)(20,45)(0)

\Gluon(100,135)(100,172){4}{3}
\Gluon(140,172)(140,135){4}{3}

\Gluon(100,28)(100,65){4}{3}
\Gluon(140,65)(140,28){4}{3}

\ArrowLine(96,158)(96,156)
\ArrowLine(144,156)(144,158)

\ArrowLine(96,44)(96,42)
\ArrowLine(144,42)(144,44)

\Text(85,157)[]{$q_1$}
\Text(165,157)[]{$q_1-q$}

\Text(85,43)[]{$q_2$}
\Text(165,43)[]{$q_2-q$}

\GCirc(120,100){40}{1}
\Text(120,100)[]{$G(\vec q_1,\vec q_2;\vec q)$}

\ArrowLine(0,10)(75,10)
\ArrowLine(165,10)(240,10)\Text(37.5,0)[]{$p_B$}
\Text(202.5,0)[]{$p_{B'}$}
\Text(120,10)[]{$\Phi_{B'B}(-\vec q_2,-\vec q)$}
\Oval(120,10)(20,45)(0)

\end{picture}
\end{center}
\caption[]{Diagrammatic representation of the elastic scattering amplitude
$A + B \rightarrow A' + B'$.} 
\end{figure}
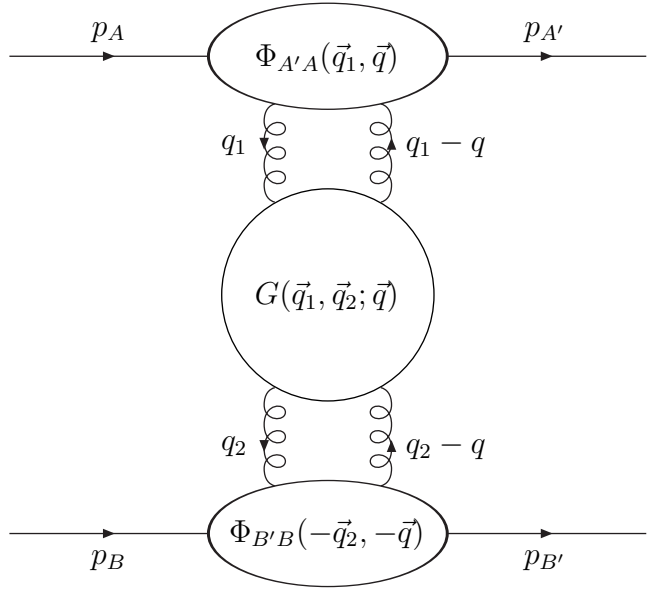    

\section{Method of calculation}
\setcounter{equation}{0}

The bootstrap conditions for the gluon Reggeization in the NLA were derived
in Ref.~\cite{FF98}. The starting point is the elastic scattering process 
$A + B \rightarrow A^\prime + B^\prime$ in the Regge kinematical region
\begin{equation}
s = (p_A + p_B)^2 = (p_A^\prime + p_B^\prime)^2 \rightarrow \infty \;,
\;\;\;t = (p_A - p_A^{\prime})^2 = (p_B^\prime - p_B)^2 \;\;\;
\mbox{--fixed}\;,
\label{RK}
\end{equation}
where $p_A$, $p_B$ and $p_A^\prime$, $p_B^\prime$ are the momenta of the
initial and final particles, respectively. We use the Sudakov decomposition 
for any vector $p$
\[
p = \beta p_1 + \alpha p_2 + p_{\perp},\;\;\;\;\;\;p_{\perp}^2
= - \vec p^{\:2}\;,
\]
where the vectors ($p_1$, $p_2$) are the light-cone basis of
the initial particle momenta plane ($p_A$, $p_B$)
\[
p_A = p_1 + \frac{m_A^2}{s} p_2 \;, \;\;\;\;\;\;p_B =
p_2 + \frac{m_B^2}{s} p_1 \;, 
\]
\[
p_A^2 = m_A^2 \;, \;\;\;\;\; p_B^2 = m_B^2 \;, \;\;\;\;\; 
p_1^2 = p_2^2 = 0 \;, \;\;\;\;\; s \approx 2( p_1 p_2) \;.
\]
In the Regge limit $ s \gg -t$, the momentum transfer is
dominated by its transverse part
\[
q = p_A - p_A^\prime \approx q_\perp\;, \;\;\;\;\;\;
t = q^2 \approx q_\perp^2 = - \vec{q}^{\:2} \;.
\]
In the case of gluon quantum numbers and negative signature in the
$t$-channel, the amplitude for this elastic process has the Regge form
\begin{equation}
({\cal A}_{8}^{(-)})_{AB}^{A^{\prime }B^{\prime }}=\Gamma _{A^{\prime }A}
^{c}\left[ \left( \frac{-s}{-t}\right) ^{j(t)}-\left( \frac{+s}{-t}\right)
^{j(t)} \right] \Gamma _{B^{\prime }B\mbox{ \ \ }}^{c} .  
\label{Ampl}
\end{equation}
Here $c$ is a color index, $\Gamma _{P^{\prime }P}^{c}$ are the 
particle-particle-Reggeon (PPR) vertices which do not depend on $s$ and
$j(t) = 1 + \omega(t)$ is the Reggeized gluon trajectory.
In the derivation of the BFKL equation, this form is assumed to be valid also
in the NLA. On the other side, the $s$-channel unitarity of the scattering
matrix leads to (see Fig.~1, where the wavy intermediate lines    
denote Reggeons) 
\[
{\cal I}{\it m}_{s}\left(({\cal A}_{{\cal R}})_{AB}^{A^{\prime}
B^{\prime }}\right) = \frac{s}{\left( 2\pi \right)^{D-2}}
\int \frac{d^{D-2}q_1}{\vec{q}_{1}^{\:2}\left( \vec{q}_{1}-\vec{q}\right)^{2}}
\int \frac{d^{D-2}q_2}{\vec{q}_{2}^{~2}\left( \vec{q}_{2}-\vec{q}\right) ^{2}}
\]
\begin{equation}
\times \sum_{\nu}\Phi _{A^\prime A}^{\left( {\cal R},\nu \right) }
\left( \vec{q}_{1},\vec{q};s_{0}\right)\int_{\delta -i\infty}^{\delta+i\infty} 
\frac{d\omega }{2\pi i}\left[ \left( \frac{s}{s_{0}}\right)^{\omega }
G_{\omega }^{\left( {\cal R}\right) }\left( \vec{q}_{1},\vec{q}_{2},\vec{q}
\right) 
\right] \Phi _{B^\prime B}^{\left( {\cal R},\nu \right) }\left( -\vec{q}_{2},
-\vec{q};s_{0}\right) \;\;,
\label{ImA}
\end{equation}
where ${\cal A}_{\cal R}$ stands for the scattering amplitude with
the representation ${\cal R}$ of the color group in the $t$-channel.
In the above equation, the index $\nu$ enumerates the states in the irreducible
representation 
${\cal R}$, $\Phi_{P^\prime P}^{\left( {\cal R},\nu \right)}$
are the impact factors and $G_{\omega }^{\left( {\cal R}\right)}$ is the Mellin 
transform of the Green function for the Reggeon-Reggeon scattering~\cite{FF98}.
Here and below we do not indicate the signature since it is defined by the symmetry
of the representation ${\cal R}$ in the product of the two octet representations.
The parameter $s_0$ is an arbitrary energy scale introduced in order to define 
the partial wave expansion of the scattering amplitudes through the Mellin
transform. The dependence on this parameter disappears in the full expressions 
for the amplitudes. The space-time dimension $D=4+2\epsilon$ is kept different from four
in order to regularize the infrared singularities.
The Green function obeys the generalized BFKL equation
\begin{equation}
\omega G_{\omega }^{\left( {\cal R}\right) }\left( \vec{q}_{1},\vec{q}_{2},
\vec{q}\right) = \vec{q}_{1}^{~2} \vec{q}_{1}^{\:\prime\:2}
\delta^{\left(D-2\right) }\left( \vec{q}_{1}-\vec{q}_{2}\right)
+\int \frac{d^{D-2} q_r }
{\vec{q}_r^{\:2} \vec{q}_r^{\:\prime \:2}}{\cal K}
^{\left( {\cal R}\right) }\left( \vec{q}_{1},\vec{q}_r;\vec{q}
\right) G_{\omega }^{\left( {\cal R}\right) }\left( \vec{q}_r,\vec{q}
_{2};\vec{q}\right) \;,  
\label{genBFKL}
\end{equation}
where ${\cal K}^{\left( {\cal R}\right)}$ is the kernel in the
NLA~\cite{FF98} and we have introduced the notation $q_i^\prime \equiv q_i - q$
(which will be used also in the following).

The two bootstrap conditions derived in Ref.~\cite{FF98} follow from the
comparison between the imaginary part of the amplitude (\ref{Ampl}) with the
imaginary part given by Eq.~(\ref{ImA}) in the case of the gluon
representation in the $t$-channel. In this paper we are interested in the 
second bootstrap
condition (Eq.~(35) in Ref.~\cite{FF98}) which includes the NLA correction
$\Phi_{A^{\prime}A}^{(8,a)(1)}$ to the octet impact factor and reads
\[
- i g \int\frac{d^{D-2}q_1}{(2\pi)^{D-1}}\frac{\vec q^{~2}}
{\vec q_1^{~2}
\vec q_1^{\:\prime\:2}}\sqrt{N}\Phi_{A^{\prime}A}^{(8,a)(1)}
(\vec q_1, \vec q; s_0)
= \Gamma_{A^{\prime}A}^{(a)(1)}\, \omega^{(1)}( - \vec q^{~2})
\]
\begin{equation}\label{boot2}
+ \frac{1}{2}\Gamma_{A^{\prime}A}^{(a)(B)}\left[ \omega^{(2)}( - \vec
q^{~2}) + \left( \omega^{(1)}( - \vec q^{~2}) \right)^2\ln\left( \frac{s_0}
{\vec q^{~2}} \right) \right],
\end{equation}
where $g$ is the gauge coupling constant, $N$ is the number of colors,
$\omega^{(1)}$ and $\omega^{(2)}$ are 
the one- and two-loop contributions to the Reggeized gluon trajectory, 
$\Gamma_{A^{\prime}A}^{(a)(B)}$ and $\Gamma_{A^{\prime}A}^{(a)(1)}$ are the 
Born and one-loop parts of the PPR effective vertex. 
 
\begin{figure}[t]
\begin{center}
{\parbox[t]{5cm}{\epsfysize 5cm \epsffile{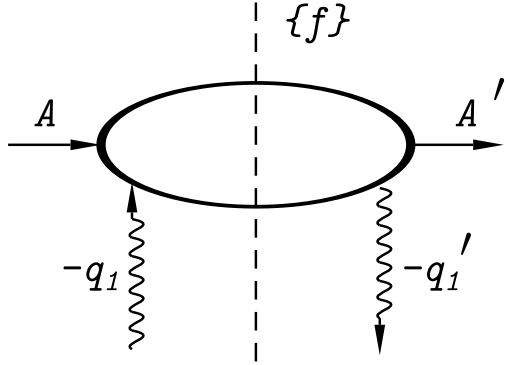}}}
\end{center}
\caption{Schematic description of the intermediate state contributions to
the impact factor.}
\label{fig}
\end{figure}

The definition of the non-forward impact factors with color state
$\nu$ of the irreducible representation ${\cal R}$ was given in 
Ref.~\cite{FF98}. This definition can be written as
\begin{equation}\label{CIF}
\Phi_{A^{\prime}A}^{({\cal R}, \nu)}(\vec q_1, \vec q; s_0) = \langle
cc^{\prime} | \hat{\cal P}_{\cal R} | \nu \rangle\Phi_{AA^{\prime}}^
{cc^{\prime}}(\vec q_1, \vec q; s_0)\;,
\end{equation}
where $\hat{\cal P}_{\cal R}$ is the projector of the two-gluon color states
in the $t$-channel on the irreducible representation ${\cal R}$
of the color group and the remaining part $\Phi_{AA^{\prime}}^{cc^{\prime}}$
is the unprojected impact factor of the particle $A$, 
\[
\Phi_{AA^{\prime}}^{cc^{\prime}}(\vec q_1, \vec q; s_0) = \left( \frac{s_0}
{\vec q_1^{~2}} \right)^{\frac{1}{2}\omega( - \vec q_1^{~2})}
\left( \frac{s_0}{\vec q_1^{\:\prime \:2}} 
\right)^{\frac{1}{2}\omega(-\vec q_1^{\:\prime \: 2})}\sum_{\{f\}}\int
\frac{ds_{AR}}{2\pi} d\rho_f \theta(s_{\Lambda} - s_{AR})
\Gamma_{\{f\}A}^c \left( \Gamma_{\{f\}A^{\prime}}^{c^{\prime}} \right)^*
\]
\begin{equation}\label{IF}
- \frac{1}{2}\int\frac{d^{D-2} q_r}{\vec q_r^{~2} \vec q_r^{\:\prime\:2}}
\Phi_{AA^{\prime}}^{c_1c_1^{\prime}(B)}(\vec q_r, \vec q)({\cal K}_r^B)_{c_1c}^
{c_1^{\prime}c^{\prime}}(\vec q_r, \vec q_1; \vec q)\ln\left(
\frac{s_{\Lambda}^2}{s_0(\vec q_r - \vec q_1)^{\:2}} \right).
\end{equation}
For brevity we do not perform here and below an explicit expansion in $\alpha_s$;
evidently this expansion is assumed and only the leading and the next-to-leading 
terms should be kept.
Here $\omega(t)$ is the Reggeized gluon trajectory which can be taken
at leading order, $\Gamma_{\{f\}A}^c$ is the effective amplitude for the 
production of the system $\{f\}$ (see Fig.~2) in the collision of the particle
$A$ with the Reggeized gluon carrying color index $c$ and momentum
\begin{equation}\label{g_mom}
- q_1 = \alpha p_B - {q_1}_{\perp},\ \ \alpha \approx \left( s_{AR} - m_A^2
+ \vec q_1^{~2} \right)/s \ll 1 \;, \;\;\; q_1^2 = - \vec{q}_1^{\:2} \;,
\end{equation}
$s_{AR}$ is the particle-Reggeon squared invariant mass and the momentum of
the other Reggeon is $q_1^\prime = q_1 - q$, with $q_1^{\prime \: 2} 
= -(\vec{q_1}-\vec{q})^{\:2}$. 
In the fragmentation region of the particles $A$ and $A^\prime$, all the
transverse 
momenta as well as the squared invariant squared mass $s_{AR}$ are of the order of 
the typical virtuality, i.e. $\vec{q}_1^{\:2}, \: s_{AR}\;\sim \vec{q}^{\:2}$. 
\newline
In Eq.~(\ref{IF}) the summation is over all the possible intermediate states
which
can be produced in the NLA and the integration is over the particle-Reggeon 
squared invariant mass and over the phase space of the produced system
\begin{equation}\label{PS}
d\rho_f = (2\pi)^D\delta^{(D)}(p_A - q_1 - \sum_{m=1}^n k_m)
\prod_{m=1}^n\frac{d^{D-1}k_m}{2\epsilon_m(2\pi)^{D-1}}\;,
\end{equation}
in the case of an $n$-particle system. 
The parameter $s_\Lambda$, which limits the integration over $s_{AR}$ in
Eq.~(\ref{IF}), is introduced in order to separate the 
contributions from multi-Regge and quasi-multi-Regge kinematics (MRK and QMRK)
and is to be considered as tending to infinity. The dependence
on this parameter disappears due to the cancellation
between the first and the second terms in the R.H.S. of Eq.~(\ref{IF}). 
The second term of Eq.~(\ref{IF}) contains the $s_0$-independent
Born contribution to the impact factor, $\Phi_{AA^{\prime}}^{cc^{\prime}(B)}$,
and the unprojected part of the non-forward BFKL kernel, connected
with real gluon production, at the Born level
\begin{equation} \label{BK}
({\cal K}_r^B)_{c_1c}^{c_1^{\prime}c^{\prime}}(\vec q_1, \vec q_2, \vec q) =
\frac{1}{2(2\pi)^{D-1}}\sum_{\lambda_G}\gamma_{c_1c}^{G(B)}(q_1, q_2)
\left( \gamma_{c_1^{\prime}c^{\prime}}^{G(B)}(q_1^{\:\prime}, q_2^{\:\prime})
\right)^*\;, 
\end{equation}
being $\gamma_{c_1c}^{G(B)}(k_1, q_1)$ the Born effective amplitude for the
production of one gluon G with helicity $\lambda_G$ in the collision of
two Reggeized gluons carrying color indices $c_1$, $c$ and momenta $q_1$, 
$-q_2$, respectively. The definitions (\ref{CIF}) and (\ref{IF}) apply also 
in the case of colorless
particle as well as in the case of charged QCD particles. Of course, the 
impact factor in the octet color representation, entering the bootstrap
condition (\ref{boot2}), makes sense only for colored particles.

In this paper we consider the non-forward quark impact factors. At the NLA
the only intermediate states $\{f\}$ which can contribute to the
quark impact factor (\ref{IF}) are one-quark and a quark-gluon system. The
second term in the R.H.S of Eq.~(\ref{IF}), which is a counterterm
for the LLA part of the first term, will be attributed to the 
quark-gluon intermediate state. In the following, we will 
determine the integral representation of the non-forward quark impact factors
in arbitrary space-time dimension $D=4+2\epsilon$ and keeping the quark
massive.
Then, after checking that the bootstrap condition (\ref{boot2}) is satisfied,
we perform the integrations for the case of massless quarks, using the 
expansion in $\epsilon$ when necessary.

\section{One-quark contribution}
\setcounter{equation}{0}

In the case of the one-quark contribution, the squared invariant mass
$s_{AR}$ is equal to the $m_A^2$, the squared mass of the colliding 
quark flavor. From the definition (\ref{IF}) one easily gets
\begin{equation}\label{1Q}
\Phi_{AA^{\prime}}^{cc^{\prime}\{Q\}}(\vec q_1, \vec q; s_0) = \left(
\frac{s_0}{\vec q_1^{\:2}} \right)^{\omega( - \vec q_1^{~2})/2}\left(
\frac{s_0}{\vec q_1^{\:\prime \:2}} \right)^{\omega( - \vec q_1^{\:\prime
\:2})/2}
\sum_{\lambda_Q}\Gamma^c_{QA}\left( \Gamma^{c^{\prime}}_{QA^{\prime}}\right)^*\;,
\end{equation}
where $\Gamma^c_{QA}$ is the quark-quark-Reggeon (QQR) effective vertex,  
which was obtained in Refs.~\cite{FFQ94,FFQ96} and has the form
\begin{equation}\label{QQR}
\Gamma^c_{QA} = g (t^c)_{QA}\left[ \delta_{\lambda_Q, \lambda_A}\left(
1+ \Gamma_{QQ}^{(+)(1)}(q_1^2) \right) + \delta_{\lambda_Q, - \lambda_A}
\Gamma_{QQ}^{(-)(1)}(q_1^2) \right]\;.
\end{equation}
Here $\Gamma_{QQ}^{(\pm)(1)}$ represent the radiative corrections to the
he\-li\-ci\-ty con\-ser\-ving and to the helicity non-con\-ser\-ving parts 
of the QQR effective interaction 
vertex, $ t^c_{QA}$ is the $(Q,A)$ matrix element of the color group generator
$t^c$ in the fundamental representation and the $\delta$-symbols
on the helicities $\lambda_Q$ and $\lambda_A$ of the quarks $A$ and $Q$ are 
defined as
\begin{equation}\label{delta+}
\delta_{\lambda_Q,\lambda_A} \equiv \overline{u}(p_Q)\frac{\not{p}_B}{s}
u(p_A)\;,
\end{equation}
\begin{eqnarray}
\delta_{\lambda_Q,-\lambda_A} &\equiv& \frac{i}{\sqrt{-(p_A-p_Q)^2}}
\overline{u}(p_Q)(\not{p}_A-\not{p}_Q)\frac{\not{p}_B}{s}u(p_A) \nonumber \\
\label{delta-}
&=& \frac{i}{\sqrt{-(p_A-p_Q)^2}}
\overline{u}(p_Q)\left(1- 2m_A \frac{\not{p}_B}{s}\right)u(p_A) \;.
\end{eqnarray}
The convolution in Eq.~(\ref{1Q}) can be easily calculated and gives
\[
\Phi_{AA^{\prime}}^{cc^{\prime}\{Q\}}(\vec q_1, \vec q; s_0) = 
g^2 (t^{c^\prime}t^c)_{A^\prime A}\left[ \delta_{\lambda_{A^\prime}, \lambda_A}
\left( 1 
+ \frac{1}{2}\omega^{(1)}(-\vec{q}_1^{\:2})\ln\left(\frac{s_0}
{\vec{q}_1^{\:2}}\right) 
\right. \right.
\]
\begin{equation}\label{1Qconv}
\left.
+ \frac{1}{2}\omega^{(1)}(-\vec{q}_1^{\:\prime \:2})\ln\left(\frac{s_0}
{\vec{q}_1^{\:\prime \:2}}\right) 
+ \Gamma_{QQ}^{(+)(1)}(-\vec{q}_1^{\:2})
+ \Gamma_{QQ}^{(+)(1)}(-\vec{q}_1^{\:\prime \:2}) \right)
\end{equation}
\[
\left.
+ \Gamma_{QQ}^{(-)(1)}(-\vec{q}_1^{\:2})  \frac{i}{\sqrt{\vec{q}_1^{\:2}}}
\overline{u}(p_{A^\prime})\frac{ \not{q}_{1_\perp}\not{p}_B }{s}u(p_A)  
+ \Gamma_{QQ}^{(-)(1)}(-\vec{q}_1^{\:\prime \:2})  
\frac{i}{\sqrt{\vec{q}_1^{\:\prime \:2}}} \overline{u}(p_{A^\prime})
\frac{\not{p}_B \not{q}^\prime_{1_\perp}}{s}u(p_A)\right]\;.
\]
Here the one-loop contribution to the Reggeized gluon trajectory
$\omega^{(1)}(-\vec{v}^{\:2})$ is given by~\cite{Lip76}
\begin{equation}\label{omega1}
\omega^{(1)}(- \vec v^{\:2}) = - \frac{g^2 N}{2}\int\frac{d^{D-2}k}{(2\pi)
^{D-1}}\:\frac{\vec v^{\:2}}{\vec k^{\:2}(\vec k - \vec v)^2} =  - \frac{g^2 N}
{(4\pi)^{2 + \epsilon}}\,\Gamma(1-\epsilon)\,\frac{[\Gamma(\epsilon)]^2}
{\Gamma(2\epsilon)}(\vec v^{\:2})^{\epsilon}\;.
\end{equation}
The radiative corrections $\Gamma_{QQ}^{(+)(1)}$ have the form~\cite{FFQ96}:
\begin{equation}\label{Gamma+}
\Gamma_{QQ}^{(+)(1)}(- \vec v^{\:2}) = a_f^{(+)}(-\vec v^{\:2}) 
+ a_Q^{(+)}(-\vec v^{\:2},m_A^2) + a_g^{(+)}(-\vec v^{\:2})
+ \delta_g^{(+)}(-\vec v^{\:2},m_A^2)
 \;,
\end{equation}
with
\begin{equation}\label{a_f+}
a_f^{(+)}(-\vec v^{\:2}) = - g^2 \frac{2}{(4\pi)^{2+\epsilon}} \Gamma
(-\epsilon)
\sum_f \int_0^1 dx \: x(1-x)\:[m_f^2+x(1-x)\vec v^{\:2}]^\epsilon\;,
\end{equation}
\[
a_Q^{(+)}(-\vec v^{\:2},m_A^2) = \frac{g^2}{2N} \frac{\Gamma(-\epsilon)}
{(4\pi)^{2+\epsilon}} 
\]
\begin{equation}\label{a_Q+}
\times \left\{ \int_0^1 \frac{dx}{[m_A^2+x(1-x)
\vec v^{\: 2}]^{1-\epsilon}}
\left[ -\vec v^{\:2} \left(\frac{1}{1+2\epsilon} + \frac{\epsilon}{2}\right)
- \frac{2 m_A^2}{1+2\epsilon}\right]+\frac{2}{1+2\epsilon} (m_A^2)^\epsilon
\right\}
\;,
\end{equation}
\[
a_g^{(+)}(-\vec v^{\:2}) = g^2 N \,
\frac{\Gamma(-\epsilon)}{(4\pi)^{2+\epsilon}}\frac{[\Gamma(1+\epsilon)]^2}
{\Gamma(1+2\epsilon)}(\vec v^{\:2})^\epsilon 
\biggl\{\psi(1-\epsilon) - 2 \psi(\epsilon) + \psi(1) 
\biggr.
\]
\begin{equation}\label{a_g+}
\left.
+ \frac{1}{1+2\epsilon}\left[\frac{1}{4(3+2\epsilon)} - \frac{1}{\epsilon}
-\frac{7}{4}\right]\right\}\;, 
\end{equation}
\[
\delta_g^{(+)}(-\vec v^{\:2},m_A^2)= 
\frac{g^2 N}{(4\pi)^{2+\epsilon}} \Gamma(1-\epsilon) \left\{ \int_0^1 dx_1\: 
\int_0^1 dx_2 \; \theta(1-x_1-x_2) \right.
\]
\[
\times \left[ \frac{-\vec v^{\:2} (1-x_1)}{x_1}\left(1-x_1+\frac{1+\epsilon}
{2}x_1^2
\right)\left(\frac{1}{[m_A^2 x_1^2 + x_2 (1-x_1-x_2)\vec v^{\:2} ]
^{1-\epsilon}}\right.\right.
\]
\begin{equation}\label{d_g+}
\left. \left. \left. \left.
- \frac{1}{[ x_2 (1-x_1-x_2)\vec v^{\:2}]^{1-\epsilon}}\right) \right.
- \frac{2m_A^2x_1}{[m_A^2 x_1^2 + x_2 (1-x_1-x_2) \vec v^{\:2}]^{1-\epsilon}}
\right] + \frac{(m_A^2)^\epsilon}{\epsilon(1+2\epsilon)} \right\}\;.
\end{equation}
The expression for $\Gamma_{QQ}^{(-)(1)}$ can be found in Ref.~\cite{FFQ94}:
\begin{equation}\label{Gamma-}
\Gamma_{QQ}^{(-)(1)}(- \vec v^{\:2}) = a_g^{(-)}(-\vec v^{\:2}) 
+ a_Q^{(-)}(-\vec v^{\:2},m_A^2)
\;,
\end{equation}
with
\begin{equation}\label{a_f-}
a_g^{(-)}(-\vec v^{\:2}) = 
-i\sqrt{\vec v^{\:2}}\, m_A \, \frac{g^2}{2N} \, \frac{\Gamma(1-\epsilon)} 
{(4\pi)^{1+\epsilon}} \frac{1-2\epsilon}{1+2\epsilon} 
\int_0^1\frac{dx}{[m_A^2 + x(1-x) \vec v^{\:2}]^{1-\epsilon}}
\end{equation}
and
\begin{equation}\label{a_Q-}
a_Q^{(-)}(-\vec v^{\:2},m_A^2) = -i\sqrt{\vec v^{\:2}} \, m_A \, g^2 N \, 
\frac{\Gamma(1-\epsilon)}{(4\pi)^{1+\epsilon}}\int_0^1dx_1 \:\int_0^1 dx_2\:
\frac{\theta(1-x_1-x_2)[x_1-(1+\epsilon)x_1^2]}
{[m_A^2 x_1^2 + x_2 (1-x_1-x_2)\vec v^{\:2}]^{1-\epsilon}}\;.
\end{equation}
In the above equations, $\Gamma(x)$ is the Euler $\Gamma$-function and
$\psi(x)$ 
its logarithmic derivative. We observe, moreover, that $
\delta_g^{(+)}$, $a_g^{(-)}$ and $a_Q^{(-)}$ vanish
in the zero quark mass limit.

\section{Quark-gluon contribution}
\setcounter{equation}{0}

In this Section we calculate the NLA contribution to the quark impact factor
from the quark-gluon production in the fragmentation region
\begin{equation}\label{QG-IF}
\Phi_{AA^{\prime}}^{cc^{\prime}(1)\{QG\}}(\vec q_1, \vec q) = 
\sum_{\stackrel{\lambda_Q, \lambda_G}{Q, g}}\int\frac{ds_{AR}}{(2\pi)}
d\rho_{\{QG\}} \: \theta(s_\Lambda-s_{AR})\Gamma^c_{\{QG\}A}(q_1)\left( 
\Gamma^{c^{\prime}}_{\{QG\}A^{\prime}}(q_1^{\prime}) \right)^* \;,
\end{equation}
where the summation is performed over the helicities $\lambda_G$ and
$\lambda_Q$
and over the color indices $g$ and $Q$ of the produced gluon and quark, having
momenta $k_1$ and $k_2$, respectively (although we use the notation $g$ also
for the coupling constant, this can not be misleading here). We remind that
the above expression must be
completed including the counterterm, i.e. the second term in the R.H.S. of 
Eq.~(\ref{IF}). We will consider this term at the end of the Section.
Introducing the Sudakov representation for the momenta of the produced
particles,
\begin{equation}\label{k1}
k_1 = \beta_1 p_1 + \frac{\vec k_1^{\:2}}{s\beta_1} p_2 +
k_{1_\perp}\;,\;\;\; k_1^2 = 0\;,
\end{equation}
\begin{equation}\label{k2}
k_2 = \beta_2 p_1 + \frac{m_A^2+\vec k_2^{\:2}}{s\beta_2} p_2 +
k_{2_\perp}\;,\;\;\; k_2^2 = m_A^2\;.
\end{equation}
we have
\begin{equation}\label{s_AR}
s_{AR} = (k_1 + k_2)^2 = \frac{\beta_1(\beta_1+\beta_2) m_A^2 + 
(\vec k_1\beta_2 - \vec k_2\beta_1)^2}{\beta_1\beta_2}\;,
\end{equation}
\begin{equation}\label{ds_AR drho}
\frac{ds_{AR}}{2\pi}d\rho_{\{QG\}} = \delta(1 - \beta_1 - \beta_2)\:
\delta^{(D-2)}\left( (k_1 + k_2 + q_1)_{\perp} \right)\frac{d\beta_1d\beta_2}
{\beta_1\beta_2}\:\frac{d^{D-2}k_1 \, d^{D-2}k_2}{2(2\pi)^{D-1}}.
\end{equation}
In the last of these equations we have used also Eq.~(\ref{g_mom}).

The amplitude for quark-gluon production in the quark-Reggeon collision
$\Gamma^c_{\{QG\}A}$ was obtained in Ref.~\cite{FFQ96} using for convenience
the the gauge $(e p_2)=0$, although all the calculations could be performed in a 
gauge invariant way. It can be written in the following form:
\begin{equation}\label{GammaQG}
\Gamma^c_{\{QG\}A} = g^2 \biggl[(t^c t^g)_{QA} {\cal A}_1 -  
(t^g t^c)_{QA} {\cal A}_2 \biggr]\;,
\end{equation}
with
\begin{equation}\label{A_1}
{\cal A}_1 = \overline{u}(k_2) \biggl[ L(k_{1_\perp})-L(-k_{2_\perp})\biggr]
\frac{\not{p}_B}{s} u(p_A)
\end{equation}
and
\begin{equation}\label{A_2}
{\cal A}_2 =  \overline{u}(k_2) \biggl[ L(\beta_2 k_{1_\perp} - \beta_1 k_{2_\perp})
-L(-k_{2_\perp})\biggr]\frac{\not{p}_B}{s} u(p_A)\;,
\end{equation}
where
\begin{equation}
L(k_\perp) = L_\mu(k_\perp) e^{* \, \mu}_\perp \;, \;\;\;\;\; 
L^\mu(k_\perp) = \frac{\gamma_\perp^\mu(m_A \beta_1^2 - \beta_1\not k_\perp)
+2 k_\perp^\mu}{m_A^2 \beta_1^2-k_\perp^2}\;,
\end{equation}
being $e^\mu$ the gluon polarization vector and $\gamma_\perp^\mu$ the
transverse component of the Dirac $\gamma^\mu$, defined as
\begin{equation}
\gamma_\perp^\mu = g_\perp^{\mu \nu} \gamma_\nu\;, \;\;\;\;\;
g^{\mu \nu} = \frac{p_A^\mu p_B^\nu + p_A^\nu p_B^\mu}{(p_A p_B)} 
+ g_\perp^{\mu \nu} \;.
\end{equation}

The amplitude $\Gamma^{c^\prime}_{\{QG\}A^\prime}$ has the same form of
the amplitude $\Gamma^c_{\{QG\}A}$, except that the Sudakov basis
$(p_1, p_2)$ must be replaced by the {\it primed} one
$(p_{1^\prime}, p_{2^\prime})$,
i.e. by the light-cone basis of the final particle momenta plane 
($p_{A^\prime}$, $p_{B^\prime}$). This leads to
\begin{equation}\label{Gamma'QG}
\Gamma^{c^\prime}_{\{QG\}A^\prime} = g^2 \biggl[(t^{c^\prime} 
t^g)_{QA^\prime} {\cal A}_1^\prime - (t^g t^{c^\prime})_{QA^\prime} 
{\cal A}_2^\prime \biggr]\;,
\end{equation}
with
\begin{equation}\label{A'_1}
{\cal A}_1^\prime = \overline{u}(k_2)\biggl[ L(k_{1_\perp}+\beta_1 q_\perp)
-L(-k_{2_\perp}-\beta_2 q_\perp)] \biggr] \frac{\not{p}_B}{s} u(p_{A^\prime})
\end{equation}
and
\begin{equation}\label{A'_2}
{\cal A}_2^\prime = \overline{u}(k_2)\biggl[ L(\beta_2 k_{1_\perp}-\beta_1 
k_{2_\perp}) -L(-k_{2_\perp}-\beta_2 q_\perp)\biggr]
\frac{\not{p}_B}{s} u(p_{A^\prime})\;.
\end{equation}

We can now calculate the convolution appearing in the integrand of 
Eq.~(\ref{QG-IF}):
\[
\sum_{\stackrel{\lambda_Q, \lambda_G}{Q,g}}\Gamma^c_{\{QG\}A}
\left(\Gamma^{c^{\prime}}_{\{QG\}A^{\prime}}\right)^*
=
\]
\[
g^4 \,\biggl[ (t^g t^{c^\prime} t^c t^g)_{A^\prime A} \sum_{\lambda_Q,
\lambda_G}
{\cal A}_1 ({\cal A}_1^\prime)^* 
- (t^{c^\prime} t^g t^c t^g)_{A^\prime A} \sum_{\lambda_Q, \lambda_G}
{\cal A}_1 ({\cal A}_2^\prime)^* \biggr.
\]
\[
\biggl.
- (t^g t^{c^\prime} t^g t^c)_{A^\prime A} \sum_{\lambda_Q, \lambda_G}
{\cal A}_2 ({\cal A}_1^\prime)^* 
+ (t^{c^\prime} t^g t^g t^c)_{A^\prime A} \sum_{\lambda_Q, \lambda_G}
{\cal A}_2 ({\cal A}_2^\prime)^* \biggr] 
\]
\[
= \frac{g^4}{4} \delta_{A^\prime A} \delta^{c c^\prime} \sum_{\lambda_Q,
\lambda_G}
{\cal A}_1 ({\cal A}_1^\prime)^*
+\frac{g^4}{2N} (t^{c^\prime}t^c)_{A^\prime A} \sum_{\lambda_Q, \lambda_G}
\biggl[ N^2\, {\cal A}_2 ({\cal A}_2^\prime)^*
\biggr.
\]
\begin{equation}\label{QGconv}
\biggl.
+ {\cal A}_1 ({\cal A}_2^\prime)^* + {\cal A}_2 ({\cal A}_1^\prime)^*
 - {\cal A}_1 ({\cal A}_1^\prime)^* - {\cal A}_2 ({\cal A}_2^\prime)^* 
\biggr]\;.
\end{equation}
In the last equality the following relations have been used:
\[
(t^g t^g)_{A^\prime A} = \frac{N^2-1}{2N} \delta_{A^\prime A}\;, \;\;\;
(t^g t^c t^g)_{A^\prime A} = -\frac{1}{2N} (t^c)_{A^\prime A}\;, 
\]
\begin{equation}
(t^g t^{c^\prime} t^c t^g)_{A^\prime A} = \frac{1}{4} \delta_{A^\prime A}
\delta^{c c^\prime} - \frac{1}{2N} (t^{c^\prime} t^c)_{A^\prime A} \;.
\end{equation}
Using the definitions of the quantities ${\cal A}_{1,2}$ and 
${\cal A}_{1,2}^\prime$ given in Eqs.~(\ref{A_1}), (\ref{A_2}), (\ref{A'_1})
and (\ref{A'_2}) and recalling that 
\begin{equation}
\sum_{\lambda} e_\perp^\mu (k,\lambda) e_\perp^{*\,\nu} (k,\lambda) =
- g_\perp^{\mu \nu}\;,
\end{equation}
it is easy to find 
\[
\sum_{\lambda_Q, \lambda_G} {\cal A}_2 ({\cal A}_2^\prime)^* \equiv - A_Q = 
- \overline{u}(p_{A^\prime}) \frac{\not{p}_B}{s} 
\biggl[ \overline{L}_\mu(\beta_2 k_{1_\perp}-\beta_1 k_{2_\perp}) - 
\overline{L}_\mu(- k_{2_\perp}-\beta_2 q_\perp) \biggr] 
\]
\begin{equation}\label{A_Q}
\times (\not{k}_2+m_A) \biggl[ L^\mu(\beta_2 k_{1_\perp}-\beta_1 k_{2_\perp})
- L^\mu(- k_{2_\perp}) \biggr] \frac{\not{p}_B}{s} u(p_A)\;, 
\end{equation}
\[
\sum_{\lambda_Q, \lambda_G} \biggl[ {\cal A}_1 ({\cal A}_2^\prime)^* 
+ {\cal A}_2 ({\cal A}_1^\prime)^* - {\cal A}_1 ({\cal A}_1^\prime)^* 
- {\cal A}_2 ({\cal A}_2^\prime)^* \biggr] \equiv  B_Q = 
\]
\[
\overline{u}(p_{A^\prime}) \frac{\not{p}_B}{s} 
\biggl[ \overline{L}_\mu(k_{1_\perp}+\beta_1 q_\perp) - 
\overline{L}_\mu(\beta_2 k_{1_\perp}-\beta_1 k_{2_\perp}) \biggr] 
\]
\begin{equation}\label{B_Q}
\times (\not{k}_2+m_A) \biggl[ L^\mu(k_{1_\perp}) - 
L^\mu(\beta_2 k_{1_\perp} - \beta_1 k_{2_\perp}) \biggr] \frac{\not{p}_B}{s} 
u(p_A)\;,
\end{equation}
\[
\sum_{\lambda_Q, \lambda_G} {\cal A}_1 ({\cal A}_1^\prime)^* \equiv - C_Q = 
- \overline{u}(p_{A^\prime}) \frac{\not{p}_B}{s} 
\biggl[ \overline{L}_\mu(k_{1_\perp}+\beta_1 q_\perp) - 
\overline{L}_\mu(-k_{2_\perp}-\beta_2 q_\perp) \biggr] 
\]
\begin{equation}\label{C_Q}
\times (\not{k}_2+m_A) \biggl[ L^\mu(k_{1_\perp}) - L^\mu(- k_{2_\perp})
\biggr]\frac{\not{p}_B}{s} u(p_A)\;, 
\end{equation}
where 
\begin{equation}
\overline{L}^\mu(k_\perp) = \frac{(m_A \beta_1^2 - \beta_1 \not{k}_\perp)
\gamma_\perp^\mu + 2 k_\perp^\mu}{m_A^2\beta_1^2-k_\perp^2}\;.
\end{equation}
The above expressions for $A_Q$, $B_Q$ and $C_Q$ can be put in the following
form

\[
A_Q = A_Q^{(+)} + A_Q^{(-)} \;,
\]
\[
A_Q^{(+)} = \beta_2^2 \biggl\{ [ 2(1-\beta_1)^2+(1+\epsilon)\beta_1^2(1-
\beta_1)]\: d(k_{2_\perp}) d(k_{2_\perp}+\beta_2 q_\perp) d(\beta_2
k_{1_\perp}-\beta_1 k_{2_\perp})\biggr.
\]
\begin{equation}\label{A+}
\times \left[\frac{\vec q^{\:2}}{d(\beta_2 k_{1_\perp}-\beta_1 k_{2_\perp})}
- \frac{\vec q_1^{\:2}}{d(k_{2_\perp}+\beta_2 q_\perp)}
- \frac{\vec q_1^{\:\prime\:2}}{d(k_{2_\perp})} \right]
+4 m_A^2 \beta_1^2
\end{equation}
\[
\biggl.
\times \biggl[ d(\beta_2 k_{1_\perp}-\beta_1 
k_{2_\perp}) - d(k_{2_\perp}+\beta_2 q_\perp)\biggr]
\biggl[ d(\beta_2 k_{1_\perp}-\beta_1 k_{2_\perp}) - d(k_{2_\perp})\biggr]
\biggr\}\delta_{\lambda_{A^\prime},\lambda_A} \;,
\]
\[
A_Q^{(-)} = 2 m_A \beta_1^2 \beta_2 (1-\beta_1) [(1+\epsilon)\beta_1 - 1]
\overline u (p_{A^\prime}) 
\biggl[ \not{q}_\perp \: d(k_{2_\perp}) d(k_{2_\perp}+\beta_2 q_\perp)
\biggr.
\]
\begin{equation}\label{A-}
\biggl.
- \not{q}_{1_\perp} \: d(k_{2_\perp}) d(\beta_2 k_{1_\perp}-\beta_1
k_{2_\perp})
+ \not{q}_{1_\perp}^\prime \: d(k_{2_\perp}+\beta_2 q_\perp) 
d(\beta_2 k_{1_\perp}-\beta_1 k_{2_\perp}) \biggr] \frac{\not{p}_B}{s}u(p_A)
\; ;
\end{equation}

\[
B_Q = B_Q^{(+)} + B_Q^{(-)} \;,
\]
\[
B_Q^{(+)} = \beta_1^2 \beta_2 \biggl\{ [ 2\beta_2 +(1+\epsilon)\beta_1^2]
\: d(k_{1_\perp}) d(k_{1_\perp}+\beta_1 q_\perp) d(\beta_2 k_{1_\perp}-\beta_1 
k_{2_\perp})
\biggr. 
\]
\begin{equation}\label{B+}
\times \left[\frac{\vec q^{\:2}}{d(\beta_2 k_{1_\perp}-\beta_1 k_{2_\perp})}
- \frac{\vec q_1^{\:2}}{d(k_{1_\perp}+\beta_1 q_\perp)}
- \frac{\vec q_1^{\:\prime\:2}}{d(k_{1_\perp})} \right]
+4 m_A^2 \beta_2
\end{equation}
\[
\biggl.
\times \biggl[ d(\beta_2 k_{1_\perp}-\beta_1 k_{2_\perp}) 
- d(k_{1_\perp}+\beta_1 q_\perp)\biggr]
\biggl[ d(\beta_2 k_{1_\perp}-\beta_1 k_{2_\perp}) - d(k_{1_\perp})\biggr]
\biggr\}\delta_{\lambda_{A^\prime},\lambda_A} \;,
\]
\[
B_Q^{(-)} = 2 m_A \beta_1^3 \beta_2 [(1+\epsilon)\beta_1 - 1]
\overline u (p_{A^\prime}) 
\biggl[ - \not{q}_\perp \: d(k_{1_\perp}) d(k_{1_\perp}+\beta_1 q_\perp)
\biggr.
\]
\begin{equation}\label{B-}
\biggl.
+ \not{q}_{1_\perp} \: d(k_{1_\perp}) d(\beta_2 k_{1_\perp}-\beta_1
k_{2_\perp})
- \not{q}_{1_\perp}^\prime \: d(k_{1_\perp}+\beta_1 q_\perp) 
d(\beta_2 k_{1_\perp}-\beta_1 k_{2_\perp}) \biggr] \frac{\not{p}_B}{s}u(p_A)
\; ;
\end{equation}

\[
C_Q = C_Q^{(+)} + C_Q^{(-)} \;,
\]
\[
C_Q^{(+)} = \beta_2 \biggl\{ [ 2\beta_2 +(1+\epsilon)\beta_1^2]
\biggl[ \beta_1^2 \vec q^{\:2} \: d(k_{1_\perp}) d(k_{1_\perp}+\beta_1 q_\perp)
+ \beta_2^2 \vec q^{\:2}\: d(k_{2_\perp}) d(k_{2_\perp}+\beta_2 q_\perp)
\biggl. \biggr.
\]
\begin{equation} \label{C+}
\biggl.
- (\vec q_1 - \beta_2 \vec q)^2 \: d(k_{1_\perp}) d(k_{2_\perp}+\beta_2
q_\perp) - (\vec q_1 - \beta_1 \vec q)^2 \: d(k_{2_\perp}) d(k_{1_\perp}
+\beta_1 q_\perp)\biggr]
\end{equation}
\[
\biggl.
+ 4 m_A^2 \beta_1^2 \beta_2 \biggl[ d(k_{1_\perp}+\beta_1 q_\perp) -
d(k_{2_\perp}+\beta_2 q_\perp)\biggr] \biggl[ d(k_{1_\perp}) - d(k_{2_\perp})
\biggr]\biggr\}\delta_{\lambda_{A^\prime},\lambda_A} \;,
\]
\[
C_Q^{(-)} = 2 m_A \beta_1^2 \beta_2 [(1+\epsilon)\beta_1 - 1]
\overline u (p_{A^\prime}) 
\biggl[ - \beta_1\!\not{q}_\perp \: d(k_{1_\perp}) d(k_{1_\perp}+\beta_1
q_\perp)
+ \beta_2\!\not{q}_\perp \: d(k_{2_\perp}) d(k_{2_\perp}+\beta_2 q_\perp)
\biggr.
\]
\begin{equation}\label{C-}
- (\not{q}_{1_\perp}-\beta_1\!\not{q}_\perp) \: d(k_{2_\perp})d(k_{1_\perp}
+\beta_1 q_\perp) 
+ (\not{q}_{1_\perp}-\beta_2\!\not{q}_\perp) \: d(k_{1_\perp})d(k_{2_\perp}
+\beta_2 q_\perp) 
\biggr] \frac{\not{p}_B}{s}u(p_A) \;;
\end{equation}
where we have used $d(l)=1/(m_A^2\beta_1^2-l^2)$ and have explicitly separated 
the helicity conserving terms (labeled by $(+)$) from the terms which,
after integration, give the helicity non-conserving contribution to the
bootstrap condition (\ref{boot2}) (labeled by $(-)$).

Putting together the above results and using Eq.~(\ref{ds_AR drho}),
the quark-gluon production contribution to the impact factor can be written as
\[
\Phi_{AA^{\prime}}^{cc^{\prime}(1)\{QG\}}(\vec q_1, \vec q) = 
\]
\begin{equation}\label{QG-IF-f}
g^4 \, \int \frac{d^{D-2}k_1} {(2\pi)^{D-1}} 
\int_{\beta_0}^1\frac{d\beta}{2\beta(1-\beta)}
\left\{ - \delta_{A^\prime A} \delta^{c c^\prime} \frac{C_Q}{4}
+ \frac{1}{2N} (t^{c^\prime} t^c)_{A^\prime A} \biggl[ -N^2 A_Q + B_Q \biggr ]
\right \} \;,
\end{equation}
where now $\beta=\beta_1$ and in the expressions for $A_Q$, $B_Q$ and $C_Q$
the 
variables $\beta_2$ and $k_{2_\perp}$ are replaced with $(1-\beta)$ and
$-q_{1_\perp}-k_{1_\perp}$, respectively. The lower limit in the 
integration over $\beta$ is $\beta_0=\vec k_1^{\:2}/s_\Lambda$ and follows from
the cut in the integration over $s_{AR}$ given by the $\theta$-function in the
Eq.~(\ref{QG-IF}). 

Let us consider first the integral
\begin{equation}\label{int-A+}
I_A^{(+)}(\vec q_1, \vec q)= \int \frac{d^{D-2}k_1}{(2\pi)^{D-1}} 
\int_{\beta_0}^1\frac{d\beta}{2\beta(1-\beta)} A_Q^{(+)} \;.
\end{equation}
It is convenient to split this integral into $I_A^{(+)}(\vec q_1, \vec q,
m_A=0)$ and the remaining part, which we call $\delta I_A^{(+)}(\vec q_1,
\vec q)$. Then, we can use Eq.~(93) of Ref.~\cite{FFQ96} to write
\begin{equation}\label{I-A-0}
I_A^{(+)}(\vec q_1, \vec q, m_A=0) = 
\int \frac{d^{D-2}k_1} {(2\pi)^{D-1}} \left[\frac{\vec q^{\:2}}
{(\vec k_1-\vec q_1)^2 (\vec k_1 - \vec q_1^{\:\prime})^2}
- \frac{\vec q_1^{\:2}}{\vec k_1^{\:2} (\vec k_1 - \vec q_1)^2}
\right.
\]
\[
\left. 
- \frac{\vec q_1^{\:\prime\:2}}{\vec k_1^{\:2} (\vec k_1 - \vec q_1^{\:\prime})
^2}\right]
\, \left[ \ln\left(\frac{s_\Lambda}{\vec k_1^{\:2}}\right) + \psi(1) - 
\psi(1+2\epsilon) - \frac {3}{4(1+2\epsilon)}\right]
\delta_{\lambda_{A^\prime},\lambda_A} \;,
\end{equation}
where we recall that $\vec q_1^{\:\prime}=\vec q_1 -\vec q$.

In the calculation of $\delta I_A^{(+)}(\vec q_1,\vec q)$ we can put 
the lower limit in the 
integration over $\beta$ equal to zero, since there is no divergence for 
$s_\Lambda \rightarrow \infty$. The result is 
\[
\delta I_A^{(+)}(\vec q_1, \vec q) = \delta_{\lambda_{A^\prime},\lambda_A} \:
\frac{\Gamma(1-\epsilon)}{(4\pi)^{2+\epsilon}}
\int_0^1 d\beta \: \frac{(1-\beta)^2}{\beta}
\int_0^1 dx \left\{ [ 2(1-\beta) + (1+\epsilon)\beta^2]  \right.
\]
\[
\times \left[ \frac{\vec q^{\:2}}{c(m_A \beta, (1-\beta) q_\perp)}
- \frac{\vec q^{\:2}}{c(0, (1-\beta) q_\perp)} 
- \frac{\vec q_1^{\:2}}{c(m_A \beta, (1-\beta) q_{1_\perp})}
+ \frac{\vec q_1^{\:2}}{c(0, (1-\beta) q_{1_\perp})} \right.
\]
\[
\left.
- \frac{\vec q_1^{\:\prime\:2}}{c(m_A \beta, (1-\beta)q^\prime_{1_\perp})}
+ \frac{\vec q_1^{\:\prime\:2}}{c(0, (1-\beta)q^\prime_{1_\perp})} \right]
+ 4 m_A^2 \frac{\beta^2}{1-\beta}\left[ \frac{1}{c(m_A \beta,0)}\right.
\]
\begin{equation}\label{dI-A+}
\left.\left.
- \frac{1}{c(m_A \beta,(1-\beta) q^\prime_{1_\perp})} 
- \frac{1}{c(m_A \beta,(1-\beta)q_{1_\perp})}
+ \frac{1}{c(m_A \beta,(1-\beta)q_\perp)}\right]\right\} \;,
\end{equation}
where we have used $c(m,l)=[m^2-l^2 x(1-x)]^{1-\epsilon}$.

We consider now the integral
\begin{equation}\label{int-A-}
I_A^{(-)}(\vec q_1, \vec q)= 
\int_0^1\frac{d\beta}{2\beta(1-\beta)}\int \frac{d^{D-2}k_1}
{(2\pi)^{D-1}} A_Q^{(-)} \;,
\end{equation}
where we have put again $\beta_0=0$ since there is no divergence for infinite
$s_\Lambda$. The result of the integration over $\vec k_1$ is
\[
I_A^{(-)}(\vec q_1, \vec q) = 
\frac{2 m_A}{(4\pi)^{2+\epsilon}}\Gamma(1-\epsilon)
\int_0^1 d\beta\: \beta(1-\beta) [ (1+\epsilon)\beta-1 ]
\int_0^1 dx \: \overline u(p_{A^\prime}) \left[ 
\frac{\not{q}_\perp}{c(m_A\beta,(1-\beta)q_\perp)} \right.
\]
\begin{equation}\label{I-A-}
\left.
- \frac{\not{q}_{1_\perp}}{c(m_A\beta,(1-\beta)q_{1_\perp})} 
+ \frac{\not{q}^\prime_{1_\perp}}{c(m_A\beta,(1-\beta)q^\prime_{1_\perp})} 
\right]\frac{\not{p}_B}{s}u(p_A)\;.
\end{equation}

The next integrals we consider are 
\begin{equation}\label{int-B}
I_B^{(\pm)}(\vec q_1, \vec q) = \int_0^1\frac{d\beta}{2\beta(1-\beta)}\int 
\frac{d^{D-2}k_1} {(2\pi)^{D-1}} B_Q^{(\pm)} \;,
\end{equation}
for which again $\beta_0$ can be put equal to zero. We have
\[
I_B^{(+)}(\vec q_1, \vec q)
= \frac{\Gamma(-\epsilon)}{(4\pi)^{2+\epsilon}} \int_0^1 dx \biggl\{ \left(
\frac{\epsilon}{2}+\frac{1}{1+2\epsilon}\right)
\left[ - \frac{\vec q^{\:2}}{c(m_A, q_\perp)}
+ \frac{\vec q_1^{\:2}}{c(m_A, q_{1_\perp})}
+ \frac{\vec q_1^{\:\prime\:2}}{c(m_A, q^\prime_{1_\perp})}\right]\biggr.
\]
\begin{equation}\label{I-B+}
\left.
-\frac{2 m_A^2}{1+2\epsilon} \left[ \frac{1}{c(m_A,0)}
- \frac{1}{c(m_A,q^\prime_{1_\perp})} 
- \frac{1}{c(m_A,q_{1_\perp})}
+ \frac{1}{c(m_A,q_\perp)} \right] \right\} 
\delta_{\lambda_{A^\prime},\lambda_A} 
\end{equation}
and
\[
I_B^{(-)}(\vec q_1, \vec q)= m_A \frac{\Gamma(1-\epsilon)}{(4\pi)^{2+\epsilon}}
\frac{2\epsilon-1}{2\epsilon+1} \int_0^1 dx \;
\overline u(p_{A^\prime}) \left[ 
\frac{- \not{q}_\perp}{c(m_A,q_\perp)} \right.
\]
\begin{equation}\label{I-B-}
\left.
+ \frac{\not{q}_{1_\perp}}{c(m_A,q_{1_\perp})} 
- \frac{\not{q}^\prime_{1_\perp}}{c(m_A,q^\prime_{1_\perp})} \right] 
\frac{\not{p}_B}{s}u(p_A)\;.
\end{equation}

Finally, we consider the integrals involving $C_Q$, i.e.
\begin{equation}\label{int-C+}
I_C^{(+)} (\vec q_1, \vec q) = \int \frac{d^{D-2}k_1} {(2\pi)^{D-1}} 
\int_{\beta_0}^1\frac{d\beta}{2\beta(1-\beta)} C_Q^{(+)} 
\end{equation}
and
\begin{equation}\label{int-C-}
I_C^{(-)}(\vec q_1, \vec q) = \int_0^1\frac{d\beta}{2\beta(1-\beta)}\int 
\frac{d^{D-2}k_1} {(2\pi)^{D-1}} C_Q^{(-)} \;,
\end{equation}
where only in the latter integration $\beta_0$ can be put equal to zero.
The integral $I_C^{(+)}(\vec q_1, \vec q)$ can be expressed in the following
form:
\[
I_C^{(+)}(\vec q_1, \vec q) \: = \: I_A^{(+)}(\vec q_1, \vec q) \: + \:
I_B^{(+)}(\vec q_1, \vec q) \: + 
\]
\[
+ \delta_{\lambda_{A^{\prime}}, \lambda_A}
\frac{\Gamma(1-\epsilon)}{(4\pi)^{2+\epsilon}} 
\int_0^1 d\beta \int_0^1 dx
\biggl\{ \frac{2(1-\beta)+(1+\epsilon)\beta^2}{\beta}
\left[ 
- \frac{(\vec q_1 - (1-\beta) \vec q)^2}{c(m_A\beta,q_{1_\perp}
-(1-\beta) q_\perp)} \right. 
\]
\[
- \frac{(\vec q_1 - \beta \vec q)^2}{c(m_A\beta,q_{1_\perp}
-\beta q_\perp)} 
+ \frac{\beta^2 \vec q_1^{\:2}}{c(m_A\beta,\beta q_{1_\perp})}
+ \frac{(1-\beta)^2 \vec q_1^{\:2}}{c(m_A\beta,(1-\beta) q_{1_\perp})}
+ \frac{(1-\beta)^2 \vec q_1^{\:\prime\:2}}{c(m_A\beta,(1-\beta) 
q_{1_\perp}^\prime)}
\]
\[
\left.
+ \frac{\beta^2 \vec q_1^{\:\prime\:2}}{c(m_A\beta,\beta 
q_{1_\perp}^\prime)}\right]
+ 4 m_A^2 \beta \left[
\frac{1}{c(m_A\beta,\beta q_{1_\perp})} 
+ \frac{1}{c(m_A\beta,(1-\beta) q_{1_\perp})} 
-\frac{1}{c(m_A\beta,q_{1_\perp}-\beta q_\perp)}\right.
\]
\begin{equation}\label{I-C+}
\biggl. \left. 
- \frac{1}{c(m_A\beta,q_{1_\perp}-(1-\beta) q_\perp)} 
+ \frac{1}{c(m_A\beta,(1-\beta) q_{1_\perp}^\prime)} 
+ \frac{1}{c(m_A\beta,\beta q_{1_\perp}^\prime)} 
-2 \, \frac{1}{c(m_A\beta,0)} \right] \biggr\}\;,
\end{equation}
while for the integral $I_C^{(-)}(\vec q_1, \vec q)$ we have 
\[
I_C^{(-)}(\vec q_1, \vec q) = 2 m_A \frac{\Gamma(1-\epsilon)}{(4\pi)
^{2+\epsilon}}\int_0^1 d\beta \: \beta[ (1+\epsilon)\beta-1] \int_0^1 dx \:
\overline u(p_{A^\prime})
\left[ -\frac{\beta \not{q}_\perp}{c(m_A\beta,\beta q_\perp)} \right.
\]
\begin{equation}\label{I-C-}
\left.
+\frac{(1-\beta) \not{q}_\perp}{c(m_A\beta,(1-\beta) q_\perp)} 
- \frac{\not{q}_{1_\perp}-\beta\!\not{q}_\perp}{c(m_A\beta,q_{1_\perp}
-\beta q_\perp)}
+ \frac{\not{q}_{1_\perp}-(1-\beta)\!\not{q}_\perp}{c(m_A\beta,q_{1_\perp}
-(1-\beta) q_\perp)}
\right] \frac{\not{p}_B}{s}u(p_A)\;.
\end{equation}

As anticipated at the beginning of this Section, we take now into account
the contribution to the impact factors from the second term in the R.H.S.
of Eq.~(\ref{IF}), i.e.
\[
- \frac{1}{2}\int\frac{d^{D-2}q_r}{\vec q_r^{~2}\vec q_r ^{\:\prime\:2}}
\Phi_{AA^{\prime}}^{c_1c_1^{\prime}(B)}(\vec q_r, \vec q)({\cal K}_r^B)_{c_1c}^
{c_1^{\prime}c^{\prime}}(\vec q_r, \vec q_1; \vec q)\ln\left(
\frac{s_{\Lambda}^2}{s_0(\vec q_r - \vec q_1)^{\:2}} \right)
\;\;\; \equiv \;\;\; \mbox{counterterm}\;.
\]
Using the expression for the unprojected quark impact factor at the Born level
(see Eq.~(\ref{1Qconv}))
\begin{equation}\label{IF-Born}
\Phi_{AA^{\prime}}^{c_1c_1^{\prime}(B)}(\vec k_1, \vec q) = 
g^2 (t^{c_1^\prime}t^{c_1})_{A^\prime A}
\delta_{\lambda_{A^\prime},\lambda_A} \;,
\end{equation}
and recalling that~\cite{BFKL}
\begin{equation}
({\cal K}_r^B)_{c_1c}^{c_1^{\prime}c^{\prime}}(\vec q_r, \vec q_1; \vec q)=
\frac{g^2}{(2\pi)^{D-1}}\: \sum_d T_{c_1c}^d (T_{c_1^\prime c^\prime}^d)^* \:
\left( \frac{\vec q_r^{\:2} \vec q_1^{\:\prime\:2}+\vec q_1^{\:2}
\vec q_r ^{\:\prime\:2}}{(\vec q_r- \vec q_1)^2}-\vec q^{\:2}\right)\;,  
\label{KernBorn}
\end{equation}
where $T^d$ are color group generators in the adjoint representation, it is 
easy to obtain the following expression for the counterterm:
\[
\mbox{counterterm} = 
- g^4 \, \left( \frac{1}{4}\delta_{A^\prime A} \delta^{c c^\prime} 
+ \frac{N}{2} (t^{c^\prime} t^c)_{A^\prime A} \right)
\int\frac{d^{D-2}k_1}{2(2\pi)^{D-1}} \ln\left(\frac{s_\Lambda^2}
{\vec k_1^{\:2} \, s_0}\right) 
\]
\begin{equation}\label{counterterm_1}
\times 
\left[ \frac{\vec q_1^{\:\prime\:2}}{\vec k_1^{\:2}(\vec k_1 - 
\vec q_1^{\:\prime})^2}
+ \frac{\vec q_1^{\:2}}{\vec k_1^{\:2}(\vec k_1 - \vec q_1)^2}
- \frac{\vec q^{\:2}}{(\vec k_1 - \vec q_1)^2(\vec k_1 - \vec q_1^{\:\prime})
^2}\right] \delta_{\lambda_{A^\prime},\lambda_A}
\end{equation}
This counterterm leads to the cancellation of the $s_\Lambda$-dependence in 
Eq.~(\ref{QG-IF-f}), which comes from the integrals 
$I_A^{(+)}(\vec q_1,\vec q,m_A=0)$ and $I_C^{(+)}(\vec q_1,\vec q)$,
as it can be easily checked comparing Eqs.~(\ref{QG-IF-f}), (\ref{I-A-0}), 
(\ref{I-C+}) and (\ref{counterterm_1}).

\section{The check of the bootstrap condition}
\setcounter{equation}{0}

We have now all the contributions needed to check the bootstrap
condition given in Eq.~(\ref{boot2}). First of all, we must consider the 
quark impact factors in the octet color representation in the $t$-channel.
According to Eq.~(\ref{CIF}), this means that the contributions to the 
unprojected quark impact factors from one-quark intermediate state, 
Eq.~(\ref{1Q}), from quark-gluon intermediate state, Eq.~(\ref{QG-IF-f}), 
and from the counterterm, Eq.~(\ref{counterterm_1}), must be contracted with 
\begin{equation}\label{octet_1}
\langle cc^{\prime} | \hat{\cal P}_8 | a \rangle = \frac{f_{acc^\prime}}
{\sqrt{N}}
\;,
\end{equation}
where $f_{abc}$ are the $SU(N)$ structure constants.
Since 
\begin{equation}\label{octet_2}
\langle cc^{\prime} | \hat{\cal P}_8 | a \rangle \: \delta^{cc^\prime} = 0\;,
\;\;\;\;\;
\langle cc^{\prime} | \hat{\cal P}_8 | a \rangle \: (t^{c^\prime} t^c)
_{A^\prime A} = -i \frac{\sqrt N}{2} (t^a)_{A^\prime A}\;,
\end{equation}
we have for the octet quark impact factor at 1-loop order the following
expression:
\[
\Phi_{A^{\prime}A}^{(8, a)(1)}(\vec q_1, \vec q; s_0) = 
-i\frac{\sqrt N}{2}(t^a)_{A^\prime A}
\left\{
g^2 \, \left[ \delta_{\lambda_{A^\prime}, \lambda_A}
\biggl( \frac{1}{2}\omega^{(1)}(-\vec{q}_1^{\:2})\ln\left(\frac{s_0}
{\vec{q}_1^{\:2}}\right) \right. \right. 
\]
\[
+ \frac{1}{2}\omega^{(1)}(-\vec{q}_1^{\:\prime \:2})\ln\left(\frac{s_0}
{\vec{q}_1^{\:\prime \:2}}\right) 
+ a_f^{(+)}(-\vec q_1^{\:2}) 
+ a_Q^{(+)}(-\vec q_1^{\:2},m_A^2) + a_g^{(+)}(-\vec q_1^{\:2})
+ \delta_g^{(+)}(-\vec q_1^{\:2},m_A^2)
\]
\[
+ a_f^{(+)}(-\vec q_1^{\:\prime\:2}) 
+ a_Q^{(+)}(-\vec q_1^{\:\prime\:2},m_A^2) 
+ a_g^{(+)}(-\vec q_1^{\:\prime\:2})
+ \delta_g^{(+)}(-\vec q_1^{\:\prime\:2},m_A^2)\biggr)
\]
\[
+ \biggl(a_g^{(-)}(-\vec q_1^{\:2}) 
+ a_Q^{(-)}(-\vec q_1^{\:2},m_A^2) \biggr) \frac{i}{\sqrt{\vec{q}_1^{\:2}}}
\overline{u}(p_{A^\prime})\frac{\not{q}_{1_\perp} \not{p}_B}{s}u(p_A)  
\]
\[
\left.
+ \biggl( a_g^{(-)}(-\vec q_1^{\:\prime\:2}) 
+ a_Q^{(-)}(-\vec q_1^{\:\prime\:2},m_A^2) \biggr)
\frac{i}{\sqrt{\vec{q}_1^{\:\prime \:2}}} \overline{u}(p_{A^\prime})
\frac{\not{p}_B \not{q}^\prime_{1_\perp}}{s}u(p_A)\right]
\]
\[
-g^4 \, \frac{N}{2} \biggl( I_A^{(+)}(\vec q_1, \vec q, m_A=0) + 
\delta I_A^{(+)}(\vec q_1, \vec q) + I_A^{(-)}(\vec q_1, \vec q) \biggr) + 
\frac{g^4}{2N} \biggl( I_B^{(+)}(\vec q_1, \vec q) + 
I_B^{(-)}(\vec q_1, \vec q) \biggr)
\]
\[
- g^4 \, \frac{N}{2} \int\frac{d^{D-2}k_1}{2(2\pi)^{D-1}} 
\ln\left(\frac{s_\Lambda^2} {\vec k_1^{\:2} \, s_0}\right) 
\]
\begin{equation}\label{8IF-int}
\left.
\times \left[ \frac{\vec q_1^{\:\prime\:2}}{\vec k_1^{\:2}
(\vec k_1 - \vec q_1^{\:\prime})^2}
+ \frac{\vec q_1^{\:2}}{\vec k_1^{\:2}(\vec k_1 - \vec q_1)^2}
- \frac{\vec q^{\:2}}{(\vec k_1 - \vec q_1^{\:\prime})^2(\vec k_1 - \vec q)^2}
\right] \delta_{\lambda_{A^\prime},\lambda_A} \right\}\;,
\end{equation}
where $a_{f,Q,g}^{(+)}$ and $\delta_g^{(+)}$ are given in Eqs.~(\ref{a_f+}),
(\ref{a_Q+}), (\ref{a_g+}) and (\ref{d_g+}), respectively, 
$a_{f,Q}^{(-)}$ in Eqs.~(\ref{a_f-}) and (\ref{a_Q-}),
$I_A^{(+)}(\vec q_1, \vec q, m_A=0)$,
$\delta I_A^{(+)}(\vec q_1, \vec q)$, $I_A^{(-)}(\vec q_1, \vec q)$ and
$I_B^{(\pm)}(\vec q_1, \vec q)$ are given in Eqs.~(\ref{I-A-0}), 
(\ref{dI-A+}), (\ref{I-A-}), (\ref{I-B+}) and (\ref{I-B-}), respectively.  
We can now proceed to check the fulfillment of the bootstrap 
condition~(\ref{boot2}), whose L.H.S. and the R.H.S. read
\[
\mbox{L.H.S.} = -g \frac{N}{2} (t^a)_{A^\prime A} 
\int\frac{d^{D-2} q_1}{(2\pi)^{D-1}}\frac{\vec q^{\:2}}
{\vec q_1^{\:2} \vec q_1^{\:\prime\:2}} \left\{ g^2 
\delta_{\lambda_{A^\prime}, \lambda_A}  \biggl( \omega^{(1)}(-\vec{q}_1^{\:2})
\ln\left(\frac{s_0}{\vec{q}_1^{\:2}}\right) \right.\biggr.
\]
\[
\biggl. 
+ 2 \biggl(a_f^{(+)}(-\vec q_1^{\:2}) 
+ a_Q^{(+)}(-\vec q_1^{\:2},m_A^2) + a_g^{(+)}(-\vec q_1^{\:2})
+ \delta_g^{(+)}(-\vec q_1^{\:2},m_A^2) \biggr) \biggr)
\]
\[
+ 2 g^2\, \biggl(a_g^{(-)}(-\vec q_1^{\:2}) 
+ a_Q^{(-)}(-\vec q_1^{\:2},m_A^2) \biggr)
\frac{i}{\sqrt{\vec{q}_1^{\:2}}}
\overline{u}(p_{A^\prime})\frac{\not{q}_{1_\perp} \not{p}_B}{s}u(p_A) 
\]
\[
- g^4 \, \frac{N}{2} \biggl( I_A^{(+)}(\vec q_1, \vec q, m_A=0) + 
\delta I_A^{(+)}(\vec q_1, \vec q) + 
I_A^{(-)}(\vec q_1, \vec q) \biggr) 
+ \frac{g^4}{2N} \biggl( I_B^{(+)}(\vec q_1, \vec q) + 
I_B^{(-)}(\vec q_1, \vec q) \biggr)
\]
\[
- g^4 \, \frac{N}{2} \int\frac{d^{D-2}k_1}{2(2\pi)^{D-1}} 
\ln\left(\frac{s_\Lambda^2} {\vec k_1^{\:2} \, s_0}\right) 
\]
\begin{equation}\label{lhs}
\left.
\times \left[ \frac{\vec q_1^{\:\prime\:2}}{\vec k_1^{\:2} 
(\vec k_1 - \vec q_1^{\:\prime})^2}
+ \frac{\vec q_1^{\:2}}{\vec k_1^{\:2}(\vec k_1 - \vec q_1)^2}
- \frac{\vec q^{\:2}}{(\vec k_1 - \vec q_1)^2 (\vec k_1 - \vec
q_1^{\:\prime})^2}\right] \delta_{\lambda_{A^\prime},\lambda_A} \right\}\;,
\end{equation}
and
\[
\mbox{R.H.S} = g (t^a)_{A^\prime A}\biggl\{
\delta_{\lambda_{A^\prime},\lambda_A} \biggl[
\frac{1}{2} \omega^{(2)(Q)}(-\vec q^{\:2}) 
+ \frac{1}{2} \omega^{(2)(G)}(-\vec q^{\:2}) 
+\frac{1}{2} (\omega^{(1)}(-\vec q^{\:2}))^2 
\ln\left(\frac{s_0}{\vec q^{\:2}}\right) \biggr. \biggr.
\]
\[
\biggl. \biggl.
+ \omega^{(1)}(-\vec q^{\:2}) \biggl(a_f^{(+)}(-\vec q^{\:2}) 
+ a_Q^{(+)}(-\vec q^{\:2},m_A^2) + a_g^{(+)}(-\vec q^{\:2})
+ \delta_g^{(+)}(-\vec q^{\:2},m_A^2) \biggr)\biggr]
\]
\begin{equation}\label{rhs}
+ \delta_{\lambda_{A^\prime},-\lambda_A} \,  \omega^{(1)}(-\vec q^{\:2})
\biggl(a_g^{(-)}(-\vec q^{\:2}) 
+ a_Q^{(-)}(-\vec q^{\:2},m_A^2) \biggr)
\biggr\}\;.
\end{equation}
In the last equation, $\omega^{(2)(Q)}(-\vec q^{\:2})$ and
$\omega^{(2)(G)}(-\vec q^{\:2})$ are the quark and gluon
contributions to the two-loop gluon trajectory~\cite{F}:
\[
\omega^{(2)(Q)}(-\vec q^{\:2}) = 
2 g^4 N \frac{\Gamma(1 - \epsilon)}{(4\pi)^{2 + \epsilon}}\frac{1}{\epsilon}
\int\frac{d^{D-2}q_1}{(2\pi)^{D-1}}\frac{\vec q^{\:2}}{\vec q_1^{\:2}
\vec q_1^{\:\prime\:2}} \sum_f\int_0^1 dx \,x(1 - x)
\]
\begin{equation}\label{omega_Q}
\times\biggl[ \left( m_f^2 + x(1 - x)\vec q^{\:2} \right)^{\epsilon} 
- 2\left( m_f^2 + x(1 - x)\vec q_1^{\:2} \right)^{\epsilon} \biggr] 
\end{equation}
and
\[
\omega^{(2)(G)}(-\vec q^{\:2}) = 
\frac{g^4 N^2}{2}\int\frac{d^{D-2}q_1}
{(2\pi)^{D-1}}\frac{d^{D-2}q_2}{(2\pi)^{D-1}}\frac{\vec q^{\:2}}{\vec q_1^{\:2}
\vec q_2^{\:2}}\left\{ \frac{\vec q^{\:2}}{2 \vec q_1^{\:\prime\:2} 
\vec q_2^{\:\prime\:2}}
\ln\left( \frac{\vec q^{\:2}}{(\vec q_1 - \vec q_2)^2} \right)\right.
\]
\[
- \frac{1}{(\vec q_1 + \vec q_2 - \vec q)^2}\ln\left( \frac{(\vec q_1 +
\vec q_2)^2}{\vec q_1^{\:2}} \right) + \biggl( \frac{- \vec q^{\:2}}{2
\vec q_1^{\:\prime\:2} \vec q_2^{\:\prime\:2}} 
+ \frac{1}{(\vec q_1 + \vec q_2 - \vec q)^2}\biggr)
\]
\begin{equation}\label{omega_G}
\left.
\times\biggl[ \frac{1}{1+2\epsilon}\left(\frac{1}{4(3 + 2\epsilon)}
-\frac{1}{\epsilon} - \frac{1}{4}\right) + 2\psi(1 + 2\epsilon) 
+ \psi(1 -\epsilon) - 2\psi(\epsilon) - \psi(1) \biggr] \right\} \;.
\end{equation}
The check of the bootstrap condition is now a matter of recognizing similar
terms in the L.H.S. and the R.H.S. written above and performing cancellations. 
We can separate this task into two parts, namely we can concentrate separately 
on helicity conserving and on helicity non-conserving terms. The first part
is not quite independent of the calculation of the two-loop correction to the
gluon trajectory~\cite{F}, since it was performed
assuming that the gluon Reggeization holds, by comparison of the $s$-channel
discontinuity dictated by the Regge form (\ref{Ampl}) with that calculated
from the unitarity. Therefore, it represents more
a test of correctness of all the calculations involved in the determination of the 
trajectory and of the impact factors. The part related to the helicity 
non-conserving terms is a quite new check of the gluon Reggeization, for the spin 
structure which is absent in the LLA. It is worthwhile to mention that the 
Reggeization in this structure is not necessary for the derivation of the BFKL 
equation. Nevertheless, it is straightforward to verify that it holds, by 
considering that the following cancellations occur between a set of terms in 
the L.H.S. and a set of terms in the R.H.S.:
\begin{itemize}
\item L.H.S.: terms with $I_B^{(-)}$ and $a_g^{(-)}$; \newline
R.H.S.: term with $a_g^{(-)}$;
\item L.H.S.: terms with $I_A^{(-)}$ and $a_Q^{(-)}$; \newline
R.H.S.: term with $a_Q^{(-)}$.
\end{itemize}
The check of the bootstrap for the helicity conserving part is not less
straightforward, although the list of cancellations to perform is longer:
\begin{itemize}
\item L.H.S.: term with $a_f^{(+)}$; \newline
R.H.S.: terms with $a_f^{(+)}$ and $\omega^{(2)(Q)}$;
\item L.H.S.: term with $a_Q^{(+)}$ and $I_B^{(+)}$; 
\newline R.H.S.: term with $a_Q^{(+)}$;
\item L.H.S.: terms with $\delta_g^{(+)}$ and $\delta I_A^{(+)}$; \newline 
R.H.S.: term with $\delta_g^{(+)}$;
\item L.H.S.: terms with $\ln (s_0/\vec q_1^{\:2})$, $a_g^{(+)}$,  
$I_A^{(+)}$ and 
$\ln (s_\Lambda^2/\vec k_1^{\:2}s_0)$ (counterterm);
\newline R.H.S.: terms with $a_g^{(+)}$, $\omega^{(2)(G)}$ and 
$\ln (s_0/\vec q^{\:2})$.
\end{itemize}
In the second cancellation for the helicity non-conserving part and in the 
third cancellation for the helicity conserving part, it has been used that
\[
\int_0^1 dx_1 \int_0^1 dx_2 \theta(1-x_1-x_2) \longrightarrow
\int_0^1 d\beta \int_0^1 dx \, (1-\beta)
\]
under the change of variables $x_1=\beta$, $x_2=(1-\beta)x$.

This completes the check of the bootstrap condition for quark impact factors.

\section{Quark impact factors in massless QCD}
\setcounter{equation}{0}

In this Section we will calculate explicitly the integrals which contribute
to the quark impact factors, restricting ourselves to the case of massless
quark,
which is acceptable for all practical applications. In this case, only the
helicity 
conserving part of the quark impact factor survives and most of the remaining
integrals can be calculated for arbitrary $\epsilon$.
We will always consider the unprojected impact factor 
$\Phi_{AA^{\prime}}^{cc^{\prime}}(\vec q_1, \vec q; s_0)$ 
defined in Eq.~(\ref{IF}). In order
to obtain the quark impact factor in a definite (octet or singlet)
color representation in the $t$-channel, it is sufficient to use 
Eq.~(\ref{CIF}), with the help of Eqs.~(\ref{octet_1}) and (\ref{octet_2}), 
for the octet case, and of 
\begin{equation}\label{singlet_1}
\langle cc^{\prime} | \hat{\cal P}_0 | 0 \rangle =
\frac{\delta_{cc^\prime}}{\sqrt{N^2-1}}\;,
\end{equation}
and
\begin{equation}\label{siglet_2}
\langle cc^{\prime} | \hat{\cal P}_0 | 0 \rangle \: \delta^{cc^\prime} =
\sqrt{N^2-1}\;, 
\;\;\;\;\;
\langle cc^{\prime} | \hat{\cal P}_0 | 0 \rangle \: (t^{c^\prime} t^c)
_{A^\prime A} = \frac{\sqrt{N^2-1}}{2N} \delta_{A^\prime A}\;,
\end{equation}
for the singlet case.

We start for completeness from the quark impact factors at the Born level,
which was already given in Eq.~(\ref{IF-Born}) and reads
\begin{equation}
\Phi_{AA^{\prime}}^{cc^{\prime}(B)}(\vec k_1, \vec q) = 
g^2 (t^{c^\prime}t^c)_{A^\prime A}
\delta_{\lambda_{A^\prime},\lambda_A} \;.
\end{equation}

At one-loop level, we must consider the contribution from one-quark and 
quark-gluon intermediate states and from the counterterm -- 
Eqs.~(\ref{1Qconv}), (\ref{QG-IF-f}) and (\ref{counterterm_1}), respectively. 
Let's start from the one-quark contribution to the 
color unprojected quark impact factor which was given in Eq.~(\ref{1Qconv}). 
In the massless quark case it becomes
\[
\Phi_{AA^{\prime}}^{cc^{\prime}(1)\{Q\}}(\vec q_1, \vec q; s_0) = 
g^2 (t^{c^\prime}t^c)_{A^\prime A}\left[ \delta_{\lambda_{A^\prime}, \lambda_A}
\left(\frac{1}{2}\omega^{(1)}(-\vec{q}_1^{\:2})\ln\left(\frac{s_0}
{\vec{q}_1^{\:2}}\right) 
\right. \right.
\]
\begin{equation}\label{1Q-IF-0}
\left.\left.
+ \frac{1}{2}\omega^{(1)}(-\vec{q}_1^{\:\prime \:2})\ln\left(\frac{s_0}
{\vec{q}_1^{\:\prime \:2}}\right) 
+ \Gamma_{QQ}^{(+)(1)}(-\vec{q}_1^{\:2})\biggr|_{m_f=0}
+ \Gamma_{QQ}^{(+)(1)}(-\vec{q}_1^{\:\prime \:2})\biggr|_{m_f=0} \right)\right]\;,
\end{equation}
where 
\begin{equation}
\Gamma_{QQ}^{(+)(1)}(- \vec v^{\:2})\biggr|_{m_f=0} = a_f^{(+)}(-\vec v^{\:2})
\biggr|_{m_f=0} + a_Q^{(+)}(-\vec v^{\:2},0) + a_g^{(+)}(-\vec v^{\:2}) \;,
\end{equation}
and $m_f$ stands for the mass of any quark flavor. The integral expressions
for $a_f^{(+)}$ and $a_Q^{(+)}$ were given in Eqs.~(\ref{a_f+}) and
(\ref{a_Q+}),
respectively, while $a_g^{(+)}$ was already given in explicit form in 
Eq.~(\ref{a_g+}). The explicit forms for $a_f^{(+)}$ and $a_Q^{(+)}$
can be easily calculated in the massless quark case giving
\begin{equation}\label{a_f+0}
a_f^{(+)}(-\vec v^{\:2})\biggr|_{m_f=0} = -\frac{g^2}{(4\pi)^{2+\epsilon}}
\, n_f\Gamma(-\epsilon)\frac{[\Gamma(1+\epsilon)]^2}{\Gamma(1+2\epsilon)}
\frac{(1+\epsilon)}{(1+2\epsilon)(3+2\epsilon)}(\vec v^{\:2})^\epsilon
\end{equation}
and
\begin{equation}\label{a_Q+0}
a_Q^{(+)}(-\vec v^{\:2},0) = -\frac{g^2}{(4\pi)^{2+\epsilon}}
\, \frac{1}{N}\Gamma(-\epsilon)\frac{[\Gamma(1+\epsilon)]^2}
{\Gamma(1+2\epsilon)}
\left[\frac{1}{\epsilon(1+2\epsilon)}+\frac{1}{2}\right](\vec v^{\:2})^\epsilon
\;.
\end{equation}
Using Eq.~(\ref{omega1}) for the one-loop gluon
trajectory and summing up all the terms in the R.H.S. of Eq.~(\ref{1Q-IF-0}),
we obtain 
\[
\Phi_{AA^{\prime}}^{cc^{\prime}(1)\{Q\}}(\vec q_1, \vec q; s_0) = 
\biggl( (t^{c^\prime}t^c)_{A^\prime A} \delta_{\lambda_{A^\prime}, \lambda_A}
\, \frac{g^4}{(4\pi)^{2+\epsilon}} \, \Gamma(-\epsilon)
\frac{[\Gamma(1+\epsilon)]^2}{\Gamma(1+2\epsilon)}\, (\vec q_1 ^{\:2})
^\epsilon 
\]
\[
\times \left\{
N \ln\left(\frac{s_0} {\vec q_1 ^{\:2}}\right)
- n_f \frac{(1+\epsilon)}{(1+2\epsilon)(3+2\epsilon)}
-\frac{1}{N} \left[ \frac{1}{\epsilon(1+2\epsilon)} + \frac{1}{2} \right]
\right.
\]
\[
\left.
+ N \left[\psi(1-\epsilon) - 2 \psi(\epsilon) + \psi(1)
+ \frac{1}{4(1+2\epsilon)(3+2\epsilon)} - \frac{1}{\epsilon(1+2\epsilon)}
- \frac{7}{4(1+2\epsilon)}\right]\right\} \biggr)
\]
\begin{equation}\label{1Q-IF-0-fin}
+ \biggl( \vec q_1 \longrightarrow \vec q_1^{\:\prime}\biggr) \; .
\end{equation}

Next, we consider the contributions to the quark impact factors from
quark-gluon intermediate state, Eq.~(\ref{QG-IF-f}), and from the 
counterterm, Eq.~(\ref{counterterm_1}). In the massless quark case, we have
\[
\Phi_{AA^{\prime}}^{cc^{\prime}(1)\{QG\}}(\vec q_1, \vec q; s_0)
+ \mbox{counterterm}  = 
g^4 \, \biggl[ - \delta_{A^\prime A} \delta^{c c^\prime} \frac{I_C^{(+)}
(\vec q_1,\vec q)}{4}  \biggr.
\]
\[
\biggl.
+ \frac{1}{2N} (t^{c^\prime} t^c)_{A^\prime A} 
\biggl( -N^2 I_A^{(+)}(\vec q_1, \vec q, m_A=0) + I_B^{(+)}(\vec q_1,\vec q)
\biggr) \biggr]
\]
\begin{equation}\label{QG-IF-0}
- g^4 \, \left( \frac{1}{4}\delta_{A^\prime A} \delta^{c c^\prime} 
+ \frac{N}{2} (t^{c^\prime} t^c)_{A^\prime A} \right)
\int\frac{d^{D-2}k_1}{2(2\pi)^{D-1}} \ln\left(\frac{s_\Lambda^2}
{\vec k_1^{\:2} \, s_0}\right) 
\end{equation}
\[
\times 
\left[ \frac{\vec q_1^{\:\prime\:2}}{\vec k_1^{\:2}(\vec k_1 - 
\vec q_1^{\:\prime})^2}
+ \frac{\vec q_1^{\:2}}{\vec k_1^{\:2}(\vec k_1 - \vec q_1)^2}
- \frac{\vec q^{\:2}}{(\vec k_1 - \vec q_1)^2 (\vec k_1 - \vec q_1^{\:\prime})
^2}\right] \delta_{\lambda_{A^\prime},\lambda_A} \;,
\]
where $I_A^{(+)}(\vec q_1,\vec q,m_A=0)$, $I_B^{(+)}(\vec q_1,\vec q)$ 
and $I_C^{(+)}(\vec q_1,\vec q)$ were given in 
Eqs.~(\ref{I-A-0}), (\ref{I-B+}) and (\ref{I-C+}), respectively.
Now, let us consider the sum of the term with $I_A^{(+)}(\vec q_1,\vec q,m_A=0)$
with the part of the counterterm having the same color structure. Leaving out
the overall factor $- g^4 N/2 \, (t^{c^\prime} t^c)_{A^\prime A}$, we have
\[
\tilde I_A^{(+)}(\vec q_1,\vec q) = I_A^{(+)}(\vec q_1, \vec q, m_A=0) + 
\int\frac{d^{D-2}k_1}{2(2\pi)^{D-1}} \ln\left(\frac{s_\Lambda^2}
{\vec k_1^{\:2} \, s_0}\right) 
\]
\begin{equation}
\times 
\left[ \frac{\vec q_1^{\:\prime\:2}}{\vec k_1^{\:2}(\vec k_1 - 
\vec q_1^{\:\prime})^2}
+ \frac{\vec q_1^{\:2}}{\vec k_1^{\:2}(\vec k_1 - \vec q_1)^2}
- \frac{\vec q^{\:2}}{(\vec k_1 - \vec q_1)^2 (\vec k_1 - \vec q_1^{\:\prime})^2}
\right] \delta_{\lambda_{A^\prime},\lambda_A} \;.
\end{equation}
It is easy to show that 
\[
\tilde I_A^{(+)}(\vec q_1,\vec q)= \biggl( \delta_{\lambda_{A^\prime},\lambda_A} \left\{ 
\frac{1}{(4\pi)^{2+\epsilon}}\, \Gamma(-\epsilon)
\frac{[\Gamma(1+\epsilon)]^2}{\Gamma(1+2\epsilon)}
\left[ - 2 (\vec q^{\:2})^\epsilon \biggl( \frac{1}{2}\ln\left(\frac{s_0}
{\vec q^{\:2}}\right) + \psi(1) - \psi(1+2\epsilon)\right. \right. \biggr.
\]
\[
\left. \biggl. 
- \frac{3}{4(1+2\epsilon)} \biggr) + (\vec q_1^{\:2})^\epsilon \biggl(
-\frac{1}{\epsilon}  - \frac{3}{1+2\epsilon}
+ 2 \psi(1-\epsilon) - 2 \psi(1+2\epsilon) + 2\psi(1) - 2 \psi(\epsilon) 
\biggr) \right.
\]
\begin{equation}\label{tilde-I-A-0}
\biggl.\left.\left. 
+ 2 (\vec q_1^{\:2})^\epsilon \ln\left(\frac{s_0}{\vec q_1^{\:2}}\right) 
- \epsilon K_1 \right]\right\}\biggr)
+ \biggl( \vec q_1 \longrightarrow \vec q_1^{\:\prime} \biggr)\;,
\end{equation}
where the cancellation of the $s_\Lambda$-dependence occurred as anticipated
in Section~4. The term $K_1$ in the R.H.S. of the above expression
stands for 
\begin{equation}\label{int-k1}
K_1 = \frac{(4\pi)^{2+\epsilon}}{4} \frac{1}{\Gamma(1-\epsilon)}
\frac{\Gamma(1+2\epsilon)}{[\Gamma(1+\epsilon)]^2} 
\int\frac{d^{D-2}k}{(2\pi)^{D-1}} \ln\left(\frac{\vec q^{\:2}}
{\vec k^{\:2}}\right) \frac{\vec q^{\:2}}{(\vec k - \vec q_1)^2
(\vec k - \vec q_1^{\:\prime})^2}\;.
\end{equation}
The integral $K_1$ can be calculated only in the $\epsilon$-expansion. Its
explicit form up to the order $\epsilon$ has been determined in the Appendix 
of Ref.~\cite{FFKP99}. Here we simply quote the result
\begin{equation}
K_1 = \frac{1}{2}(\vec q^{\:2})^\epsilon 
\biggl[ \frac{1}{\epsilon^2}\biggl( 2 - \left( \frac{\vec q_1^{\:\prime\:2}}
{\vec q^{\:2}}\right)^{\epsilon} - \left( \frac{\vec q_1^{\:2}}{\vec q^{\:2}} 
\right)^{\epsilon} \biggr) + 4\psi^{\prime\prime}(1)\epsilon + \ln\left(\frac{\vec
q_1^{\:\prime\:2}}{\vec q^{\:2}} \right)\ln\left( \frac{\vec q_1^{\:2}}
{\vec q^{\:2}} \right) \biggr]\;.
\end{equation}

Similarly, we consider now the sum of the term with $I_C^{(+)}(\vec q_1,\vec q)$
with the part of the counterterm having the same color structure. Leaving out
the overall factor $-g^4/4 \:\delta_{A^\prime A} \delta^{c c^\prime} $, we have
\[
\tilde I_C^{(+)}(\vec q_1,\vec q) = I_C^{(+)}(\vec q_1,\vec q) + 
\int\frac{d^{D-2}k_1}{2(2\pi)^{D-1}} \ln\left(\frac{s_\Lambda^2}
{\vec k_1^{\:2} \, s_0}\right) 
\]
\begin{equation}\label{tilde-I_C}
\times 
\left[ \frac{\vec q_1^{\:\prime\:2}}{\vec k_1^{\:2}(\vec k_1 - 
\vec q_1^{\:\prime})^2}
+ \frac{\vec q_1^{\:2}}{\vec k_1^{\:2}(\vec k_1 - \vec q_1)^2}
- \frac{\vec q^{\:2}}{(\vec k_1 - \vec q_1)^2 (\vec k_1 - \vec
q_1^{\:\prime})^2}\right] \delta_{\lambda_{A^\prime},\lambda_A} \;.
\end{equation}
This expression can be put in the following form:
\begin{equation}\label{tilde-I-C+}
\tilde I_C^{(+)}(\vec q_1,\vec q) = \tilde I_A^{(+)}(\vec q_1,\vec q)+ 
\frac{2 \Gamma(-\epsilon)}{(4\pi)^{2+\epsilon}}\, 
\frac{[\Gamma(1+\epsilon)]^2}{\Gamma(1+2\epsilon)}
\left[ - \left(\frac{1}{\epsilon(1+2\epsilon)} + \frac{1}{2}\right)\, 
(\vec q^{\:2})^\epsilon + K_2^\prime \right]\delta_{\lambda_{A^\prime},\lambda_A}\;,
\end{equation}
where $ K_2^\prime$ is the integral analogous to  $K_2$ from 
Ref.~\cite{FFKP99}:
\[
K_2^\prime = \int_0^1 \frac{d\beta}{\beta} \: [2(1-\beta)+(1+\epsilon)\beta^2]
\left\{\biggl[ \biggl((\beta \vec q_1 + (1-\beta) 
\vec q_1^{\:\prime})^2\biggr)^\epsilon
- \biggl((1-\beta)^2 \vec q_1^{\:\prime\:2}\biggr)^\epsilon\biggr] \right.
\]
\begin{equation}\label{int-k'2}
\left. + \biggl[ \vec q_1 \longrightarrow \vec q_1^{\:\prime}\biggr] \right\}
\;.
\end{equation}
This integral can be easily calculated in the $\epsilon$-expansion up to the
order $\epsilon$ and the result is the following:
\begin{equation}
K_2^\prime = \epsilon \left[ 1 + \frac{1}{2} \ln^2\left(\frac{\vec q_1^{\:2}}
{\vec q_1^{\:\prime\:2}}\right) - \frac{3}{2} \frac{(\vec q_1^{\:2}
- \vec q_1^{\:\prime\:2})}{\vec q^{\:2}} \ln \left(\frac{\vec q_1^{\:2}}
{\vec q_1^{\:\prime\:2}}\right) - 6 \frac{|\vec q_1| |\vec q_1^{\:\prime}|}
{\vec q^{\:2}} \theta \sin \theta + 8 \psi^{\prime}(1) - 2 \theta^2 \right],
\end{equation}
being $\theta$ the angle between $\vec q_1^{\:\prime}$ and $\vec q_1$ 
defined so that $|\theta| \leq \pi$.

The last integral we need to calculate is $I_B^{(+)}(\vec q_1, \vec q)$. 
In the massless
quark case, it can be easily found that
\begin{equation}\label{I-B+0}
I_B^{(+)}(\vec q_1, \vec q) = -\frac{2}{(4\pi)^{2+\epsilon}}\, 
\Gamma(-\epsilon)\frac{[\Gamma(1+\epsilon)]^2}{\Gamma(1+2\epsilon)}
\left[\frac{1}{\epsilon(1+2\epsilon)} + \frac{1}{2}\right]\, 
\biggl[ (\vec q^{\:2})^\epsilon - (\vec q_1^{\:2})^\epsilon -
(\vec q_1^{\:\prime\:2})^\epsilon\biggr]\delta_{\lambda_{A\prime},\lambda_A}\;.
\end{equation}

Summarizing, we can write the contribution to the quark impact factors
from quark-gluon intermediate state in the compact form 
\[
\Phi_{AA^{\prime}}^{cc^{\prime}(1)\{QG\}}(\vec q_1, \vec q, s_0) 
+ \mbox{counterterm} = 
\]
\begin{equation}\label{QG-IF-0-fin}
g^4 \, \left[ - \delta_{A^\prime A} \delta^{c c^\prime} \frac{\tilde I_C^{(+)}
(\vec q_1,\vec q)}{4} + \frac{1}{2N} (t^{c^\prime} t^c)_{A^\prime A} 
\biggl( -N^2 \tilde
I_A^{(+)}(\vec q_1,\vec q) + I_B^{(+)}(\vec q_1,\vec q) \biggr) \right]\;,
\end{equation}
with $\tilde I_A^{(+)}(\vec q_1,\vec q)$, $\tilde I_C^{(+)}(\vec q_1,\vec q)$ and 
$I_B^{(+)}(\vec q_1,\vec q)$ given in Eqs.~(\ref{tilde-I-A-0}), 
(\ref{tilde-I-C+}), (\ref{I-B+0}), respectively.

In the special case of forward scattering, the expression for the quark 
impact factor at 1-loop order greatly simplifies. We have indeed
\[
\Phi_{A^{\prime}A}^{c c^\prime (1)}(\vec q_1, 0; s_0) = 
\Phi_{AA^{\prime}}^{cc^{\prime}(1)(Q)}(\vec q_1, 0; s_0)
+ \Phi_{AA^{\prime}}^{cc^{\prime}(1)\{QG\}}(\vec q_1, 0, s_0) 
+ \mbox{counterterm}|_{t=0}\;.
\]
The contribution from the one-quark intermediate state, 
$\Phi_{AA^{\prime}}^{cc^{\prime}(1)(Q)}(\vec q_1, 0; s_0)$, 
is given in the explicit form for arbitrary $\epsilon$ 
by the R.H.S. of Eq.~(\ref{1Q-IF-0-fin}) evaluated at $t=0$.
In order to determine the contribution from the quark-gluon intermediate state, 
$\Phi_{AA^{\prime}}^{cc^{\prime}(1)\{QG\}} (\vec q_1,0, s_0)$, and
from the counterterm at $t=0$, it is necessary to calculate the integrals 
in the R.H.S. of Eq.~(\ref{QG-IF-0-fin}) at $t=0$.. This can be 
done for arbitrary $\epsilon$ yielding the following results
\[
\tilde I_A^{(+)}(\vec q_1,0) = \delta_{\lambda_{A^\prime},\lambda_A} 
\frac{2}{(4\pi)^{2+\epsilon}}\, \Gamma(-\epsilon)
\frac{[\Gamma(1+\epsilon)]^2}{\Gamma(1+2\epsilon)} (\vec q_1^{\:2})^\epsilon
\biggl[ 2 \ln\left(\frac{s_0}{\vec q_1^{\:2}}\right) \biggr.
\]
\begin{equation}
\biggl.
+ 2 \psi(1-\epsilon) - 2 \psi(1+2\epsilon) + 2\psi(1) - 2 \psi(\epsilon)
-\frac{1}{\epsilon}  - \frac{3}{1+2\epsilon} \biggr]\;,
\end{equation}
\[
\tilde I_C^{(+)}(\vec q_1,0) = \delta_{\lambda_{A^\prime},\lambda_A} 
\frac{2}{(4\pi)^{2+\epsilon}}\, \Gamma(-\epsilon)
\frac{[\Gamma(1+\epsilon)]^2}{\Gamma(1+2\epsilon)} (\vec q_1^{\:2})^\epsilon
\biggl[ 2 \ln\left(\frac{s_0}{\vec q_1^{\:2}}\right) \biggr.
\]
\begin{equation}
\biggl.
+ 2 \psi(1-\epsilon) + 2 \psi(1+2\epsilon) - 2\psi(1) - 2 \psi(\epsilon)
-\frac{1}{\epsilon}  - 3 + \epsilon \biggr]
\end{equation}
and
\begin{equation}
I_B^{(+)}(\vec q_1, 0) = \delta_{\lambda_{A^\prime},\lambda_A} 
\frac{2}{(4\pi)^{2+\epsilon}}\, \Gamma(-\epsilon)
\frac{[\Gamma(1+\epsilon)]^2}{\Gamma(1+2\epsilon)} (\vec q_1^{\:2})^\epsilon
\biggl[ 1 + \frac{2}{\epsilon(1+2\epsilon)} \biggr]\;,
\end{equation}

Let us present in the explicit form the color singlet impact factor. Performing
the projection on the color singlet with the help of the operator~(\ref{singlet_1}),
we obtain
\[
\Phi_{A^{\prime}A}^{(0)(1)}(\vec q_1, 0; s_0)=\delta_{A^\prime,A}
\delta_{\lambda_{A^\prime},\lambda_A} 
g^4 \: \frac{2 \, \Gamma(-\epsilon)}{(4\pi)^{2+\epsilon}}\,
\frac{[\Gamma(1+\epsilon)]^2}{\Gamma(1+2\epsilon)} (\vec q_1^{\:2})^\epsilon
\frac{\sqrt{N^2-1}}{2N}
\biggl[ - n_f \, \frac{1+\epsilon}{(1+2\epsilon) (3+2\epsilon)}
\]
\[
+N \biggl(- \ln\left(\frac{s_0}{\vec q_1^{\:2}}\right) \biggr.
+ \psi(1) - \psi(1-\epsilon) + \frac{3}{2} + \frac{15}{8(1+2\epsilon)}  
- \frac{1}{8(3+2\epsilon)} - \frac{\epsilon}{2}\biggr) \biggr] 
\]
\[
\approx
\delta_{A^\prime,A}
\delta_{\lambda_{A^\prime},\lambda_A} \: g^2 \: \frac{\sqrt{N^2-1}}{2N}\:
\biggl[ -g^2\, N \, \frac{\Gamma(1-\epsilon)}{(4\pi)^{2+\epsilon}}\, 
\frac{[\Gamma(\epsilon)]^2}{\Gamma(2\epsilon)} (\vec q_1^{\:2})^\epsilon\biggr] \:
\]
\begin{equation}\label{singlet-final}
\times \biggl[ - \ln\left(\frac{s_0}{\vec q_1^{\:2}}\right) + \biggl( \frac{10}{3}
- \frac{1}{3} \frac{n_f}{N}\biggr) + \epsilon \biggl(- \frac{38}{9}
+ \frac{\pi^2}{6} + \frac{5}{9}\frac{n_f}{N}\biggr) \biggr]\;.
\end{equation}

The quark impact factor in the forward case ($t=0$ and color singlet in the 
$t$-channel) was considered in Refs.~\cite{CC98,C98,CC99}. 
In Refs.~\cite{C98,CC99}
it was calculated for massless quarks  with accuracy up to terms finite in 
the $\epsilon \rightarrow 0$ limit. In this particular case our 
result~(\ref{singlet-final}) is in agreement with the corresponding result of 
Ref.~\cite{CC99}, though the comparison is not straightforward 
because of the different definitions adopted. 
For details see Ref.~\cite{FFKP99}.  

\section{Discussion}
\setcounter{equation}{0}

In this paper we have obtained an integral representation for the NLA
non-forward 
quark impact factors with singlet and octet representation in the $t$-channel 
in QCD with massive quarks for arbitrary space-time dimension $D=4+2\epsilon$.
Using this integral representation, we have explicitly verified the fulfillment
of the ``second'' bootstrap condition derived in Ref.~\cite{FF98} for the
gluon Reggeization at the NLA in perturbative QCD. This check is very 
important since the gluon Reggeization plays the most relevant role
in the derivation of the BFKL equation at the NLA. Moreover, it represents a
test
of correctness of the calculations performed so far to determine the NLA 
corrections to the BFKL equation. Subsequently, we have carried out explicitly
the 
integrations in the case of massless quarks, using the $\epsilon$-expansion
when necessary.

We finally note that throughout the paper we have used the unrenormalized 
coupling constant $g$ and the same parameter $\epsilon$ to regularize both
infrared and ultraviolet divergences. The final result for the quark impact 
factors is affected by both kind of divergences. The ultraviolet ones
are easily removed by introducing the renormalized charge in the
$\overline{MS}$ scheme
\begin{equation}
g = g(\mu)\mu^{-\epsilon}\left[ 1 + \left( \frac{11}{3} - \frac{2}{3}\frac{n_f}
{N} \right)\frac{g^2(\mu)N\Gamma(1 - \epsilon)}{2\epsilon(4\pi)^{2 + \epsilon}}
\right]\;.
\end{equation}
The infrared divergences, however, are not canceled. This is expected
for the quark impact factor, since the quark is a colored object, 
whereas impact factors of colorless particles only must be infrared 
safe. Recall that for the colorless particles the infrared safety of 
the impact factors is guaranteed~\cite{FM99} by their definition given 
in~\cite{FF98} and used in this paper.

\end{document}